\documentclass[aps,pra,groupedaddress,superscriptaddress,twocolumn,longbibliography]{revtex4-1}

\usepackage[utf8x]{inputenc}
\usepackage{color}
\usepackage{bbm} 

\usepackage{amsfonts,amsmath,amssymb,stmaryrd}

\usepackage{graphicx}
\usepackage{subfigure}  
\usepackage{bbm}
\usepackage{hyperref}
\usepackage{epsfig}
\usepackage{mathrsfs}
\usepackage{verbatim}
\usepackage{ulem}	
\usepackage{siunitx}
\usepackage{soul} 

\usepackage[all]{hypcap}
\usepackage{braket}

\newcommand{\gtwotau}{{{g}^{(2)}\left(\tau\right)}}
\newcommand{\gtwozero}{{{g}^{(2)}\left(0\right)}}
\newcommand{\gtwo}{{{g}^{(2)}}}
\newcommand{\tc}{\tau_\mathrm{cl}}
\newcommand{\Nc}{N_\mathrm{cl}}
\newcommand{\Gs}{\Gamma_\mathrm{sp}}
\newcommand{\ts}{\tau_\mathrm{sp}}
\newcommand{\Nat}{N_\mathrm{at}}
\newcommand{\Nry}{N_\mathrm{Ry}}

\newcommand{\eff}{\mathrm{eff}}

\usepackage{array}

\begin{document}
\normalem	

\title{Bistability vs. metastability in driven dissipative Rydberg gases}

\author{F. Letscher}
\affiliation{Department of Physics and research center OPTIMAS, University of Kaiserslautern, Germany}
\affiliation{Graduate School Materials Science in Mainz, Gottlieb-Daimler-Strasse 47, 67663 Kaiserslautern, Germany}

\author{O. Thomas}
\affiliation{Department of Physics and research center OPTIMAS, University of Kaiserslautern, Germany}
\affiliation{Graduate School Materials Science in Mainz, Gottlieb-Daimler-Strasse 47, 67663 Kaiserslautern, Germany}

\author{T. Niederpr\"{u}m}
\affiliation{Department of Physics and research center OPTIMAS, University of Kaiserslautern, Germany}

\author{M. Fleischhauer}
\affiliation{Department of Physics and research center OPTIMAS, University of Kaiserslautern, Germany}

\author{H. Ott}
\affiliation{Department of Physics and research center OPTIMAS, University of Kaiserslautern, Germany}

\begin{abstract}
We investigate the possibility of a bistable phase in an open many-body system. To this end we discuss the microscopic dynamics of a continuously off-resonantly driven Rydberg lattice gas in the regime of strong decoherence. Our experimental results reveal a prolongation of the temporal correlations exceeding the lifetime of a single Rydberg excitation and show strong evidence for the formation of finite-sized Rydberg excitation clusters in the steady state. We simulate the dynamics of the system using a simplified and a full many-body rate-equation model. The results are compatible with the formation of metastable states associated with a bimodal counting distribution as well as dynamic hysteresis. Yet, a scaling analysis reveals that the correlation times remain finite for all relevant system parameters, which suggests a formation of many small Rydberg clusters and finite correlation lengths of Rydberg excitations. These results constitute strong evidence against the presence of a global bistable phase previously suggested to exist in this system.
\end{abstract}

\date{\today}

\maketitle

\begin{figure}
	\centering
	\includegraphics[width=1\columnwidth]{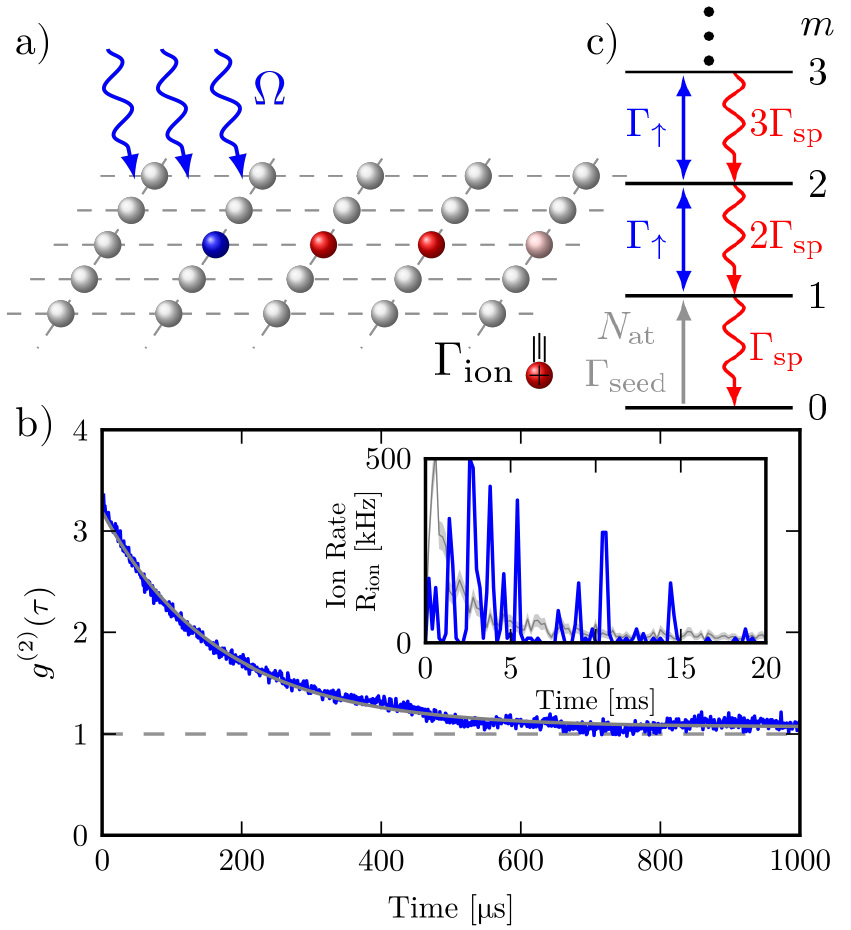}
	\caption{(a) Schematic sketch of the experimental measurement. 
	Atoms confined in a 3D optical lattice (only one plane is shown) are continuously excited to the Rydberg state with Rabi frequency $\Omega$ and detuning $\Delta > 0$. 
	The Rydberg atoms are ionized at the rate $\Gamma_\mathrm{ion}$, which is much smaller than the natural decay rate. 
	Therefore, the ions serve as a probe of the excitation dynamics. The color scheme of the atoms is chosen according to the rates in (c): Rydberg excitations are red, ground state atoms are gray and atoms, which can get facilitated are blue. (b) Pair correlation function $\gtwotau$ deduced from the ion signal ($\Omega = 2\pi\times\SI{77}{\kHz}$, $\Delta = 2\pi\times\SI{13}{\MHz}$). The correlation function shows super Poissonian fluctuations with large bunching amplitude $\gtwozero \gg 1 $  and relaxation time $\tau^{(2)} \gg \ts$ , where $\ts = \SI{20}{\us}$ is the natural lifetime of the excited Rydberg state $25\mathrm{P}_{1/2}$. Inset: measured ion signal for a single (blue) and averaged over 40 (gray) experimental realizations. The shaded area around the gray line denotes Poissonian fluctuations. (c) Effective cluster model: A single seed excitation with rate $\Nat\Gamma_\mathrm{seed}$ initializes the growth of a cluster. While a facilitation rate $\Gamma_\uparrow \gg \Gamma_\mathrm{seed}$ leads to an increase of cluster size $m$, the spontaneous decay rate $\Gs$ limits the size of a cluster.} 
	\label{fig:Singlerun}
\end{figure}

\section{Introduction}
Recent progress in experimental control offers unique possibilities to study the interplay between pure quantum mechanical systems and their coupling to reservoirs. In particular the competition between external drive and dissipation can result in interesting phases in the steady state \cite{Kessler2012, Hoening2014, Mendoza-Arenas2015, Wilson2016}. One fascinating avenue is to study the properties of these phases and the possibility of phase transitions in the context of an open quantum system \cite{Kessler2012, Lang2015, Macieszczak2016, Maghrebi2016, Weimer2015}. In addition, including strong and nonlocal interactions may allow to prepare and study strongly correlated many-body systems \cite{Ates2012, Kapit2014, Overbeck2016}. 

Rydberg atoms are particularly interesting as they provide strong and quasi-long range interactions in an inherently dissipative environment. By means of an optical drive, ground state atoms can be excited to states with high principal quantum numbers. These Rydberg states have a large polarizability resulting in a strongly interacting quantum gas. Therefore, these systems are an ideal platform to study non-equilibrium many-body physics with strong dissipation \cite{Lee2012, Carr2013, Lesanovsky2013, Hoening2014, Urvoy2015, Sibalic2016, DeMelo2016}.  

Depending on the driving scheme to the excited Rydberg state and the geometry of the atomic ensemble, Rydberg gases can exhibit very different phases. In the case of the so-called blockade regime, i.e. for resonant driving, the strong interaction shift of Rydberg states induced by an already present Rydberg excitation suppresses further excitations in a mesoscopic ensemble of surrounding atoms. This leads to the concept of a Rydberg superatom \cite{Heidemann2007, Dudin2012, Weber2015, Zeiher2015, Ebert2015} and was initially proposed to realize quantum gates \cite{Lukin2001}. Here, interesting phases with antiferromagnetic, crystalline long-range order have been predicted \cite{Hoening2014} and phases with finite-length spatial correlations have been observed \cite{Schauss2014}. In the case of the so called anti-blockade regime, i.e. for off-resonant driving, the interaction shift may be compensated by detuning the frequency of the excitation laser from resonance. Now, instead of suppressing excitations within a certain volume, further Rydberg excitations are facilitated at a specific distance from an already excited atom. This leads to the formation of Rydberg excitation clusters \cite{Gaerttner2013, Lesanovsky2014a, Urvoy2015}. 

One interesting aspect observed in recent anti-blockade experiments using a full counting analysis is a bimodal excitation number distribution \cite{Schempp2014, Malossi2014} and a hysteresis in the excitation dynamics \cite{Carr2013, DeMelo2016}.
It has been under debate whether this indicates a transition to a bistable steady state of the open many body system \cite{Ates2012, Carr2013, Hu2013, Weimer2015, Sibalic2016, Weller2016}. We will show that both, hysteresis and bimodal counting statistics, are features of metastable states. 

A metastable state manifests itself in a separation of timescales, where the relaxation time $T$ is much larger than all other timescales in the system \cite{Macieszczak2016, Rose2016}. This may lead to a hysteresis behavior upon temporal changes of system parameters on a time-scale $\tau$ (sweep time) small compared to $T$. When the sweep time becomes much larger than $T$ the hysteresis disappears. Associated with a finite value of $T$ is an upper limit for all spatial correlation lengths, $\xi\sim T$. We will argue that metastable systems with correlation length similar to the system size allow for a bimodal distribution. If $T$ increases with the system size $L$ and eventually diverges in the thermodynamic limit, the metastable states become degenerate, truly stationary states of the system. In this case the steady state is called bistable (multistable). As will be shown later bistability in an infinite, translationally invariant system implies long-range spatial order. Thus, different from metastability, the emergence of bistability can be associated with a phase transition to an ordered state.

In this paper, we experimentally investigate and theoretically model the excitation dynamics of a Rydberg lattice gas with off-resonant continuous driving and strong dissipation
(see Fig. \ref{fig:Singlerun}a). We continuously monitor the laser-induced excitation and subsequent ionization of Rydberg atoms and analyse its temporal correlations (Fig. \ref{fig:Singlerun}b) in the anti-blockade regime scanning the frequency and intensity of the excitation laser. Although there is a constant loss of atoms, our measurement technique allows us to access steady-state properties. The results indicate the formation of small clusters with a long but finite lifetime, which originate from a correlated excitation cascade. We compare the experimental results to numerical simulations based on a simplified single-cluster model (Fig. \ref{fig:Singlerun}c) and a more advanced rate equation model of extended systems. To check for the existence of bistability we perform a  finite-size extrapolation in the numerical simulations.  Our results indicate that the correlation time and length remain finite in the experimentally relevant regime amenable to a rate equation description. The bimodal character of the full counting distribution \cite{Malossi2014} disappears when the observed system size exceeds the correlation length, while a dynamic hysteresis can still be observed provided the sweep time is sufficiently small \cite{Casteels2016a}. This suggests that the Rydberg aggregate in the anti-blockade regime consists of many small independent clusters and is incompatible with a global bistable regime.

\section{Bistable and metastable steady states}
\label{sec:Bistability}

Before discussing Rydberg lattices gases 
let us start with a general introduction to bistability and metastability in open quantum systems.  
Quantum optical systems coupled to external reservoirs and subject to external drives can often be described in terms of a Lindblad master equation for the
many-body density matrix $\rho$ \cite{Lindblad1976,Gorini1976,Breuer2007}
\begin{equation}
\frac{{\rm d}}{{\rm d}t} \rho = {\cal L}\rho.
\end{equation}
The Lindbladian superoperator ${\cal L}$ is the generator of the dynamics, determined by the Hamiltonian $H$ describing the unitary evolution and
the jump operators $L_\mu$ describing the Markovian reservoir couplings
\begin{align}
{\cal L}(\rho) =& - \frac{i}{\hbar}[H,\rho] \nonumber \\ 
&+ \sum_\mu\frac{\Gamma_\mu}{2} \left(2L_\mu \rho L_\mu^\dagger -L_\mu^\dagger L_\mu \rho\, - \rho L_\mu^\dagger L_\mu\right).
\end{align}

 ${\cal L}$ has complex eigenvalues $\lambda_k$ and corresponding right eigenvectors $\rho_k$, i.e. ${\cal L}\rho_k = -\lambda_k\rho_k$. The non-negative real parts of the eigenvalues, Re$[\lambda_k]\ge 0$, determine the characteristic relaxation rates towards
a stationary state $\rho_0$ or a manifold of stationary states $\{\rho_0^{(n)}\}$ with $\lambda_0=0$. 
It should be noted that all "excited" eigenvectors $\rho_k$ with $\lambda_k\ne 0$ are no density matrices since Tr$\{\rho_k\}=0$.
If the steady state is unique, it is called stable, if two
orthogonal stationary solutions, $\rho_0^{(1)}$ and $\rho_0^{(2)}$, exist, the steady state is called bistable. In the first case any initial state will eventually evolve into
the unique steady state, in the second case the asymptotic state for large times is a mixture of the two solutions with coefficients determined
by the initial state $\rho(t=0)$:
\begin{equation}
\rho_{\rm ss} = p \rho_0^{(1)} +(1-p) \rho_0^{(2)},\qquad p = {\rm Tr}\{\check \rho_0^{(1)} \rho(t=0)\}
\label{eq:bistable-ss}
\end{equation}
Here $\check \rho_0^{(n)}$ is the left eigenvector of ${\cal L}$ corresponding to $\rho_0^{(n)}$ and Tr$\{\check\rho_0^{(n)} \rho_0^{(m)}\} = \delta_{nm}$.

In the cases we are interested in here, both the Hamiltonian $H$, as well as the jump operators $L_\mu$ are local, i.e. they can be written
as a sum of terms which act only locally. In such a case the steady state is 
generically unique if the system is {\it finite} unless the reservoir interactions are fine-tuned. General conditions for the uniqueness of steady states can be found e.g. in \cite{Davies1976,Spohn1976,Spohn1977}. 

\begin{figure}[htb]
	\begin{center}
	\includegraphics[width=0.95\columnwidth]{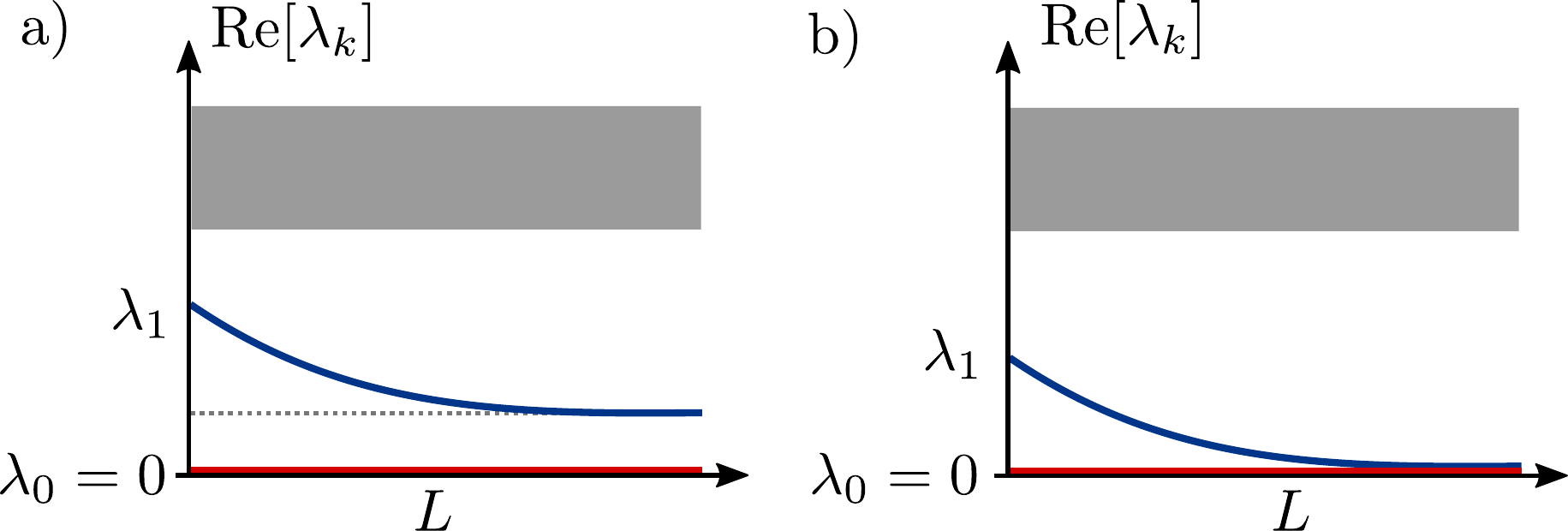}
	\end{center}
	\caption{Generic spectrum of relaxation (damping) rates as a function of system size $L$ for a) a metastable and b) a truly bistable system.} 
	\label{fig:damping-spectrum}
\end{figure}

Since experiments can be performed only in finite time it is interesting to ask if states exist that, while not being true stationary states, appear to be stable over long but finite time scales. 
These states are called {\it metastable} and they occur in systems with a separation of time scales, i.e. where there is initial relaxation into long-lived states
with the subsequent decay into the true stationary state on a much longer time scale \cite{Macieszczak2016,Rose2016}. 
Metastability is reflected in the spectrum of decay rates, Re$[\lambda_k]$, as eigenvalues with small real part, clearly separated from
all other eigenvalues, see Fig. \ref{fig:damping-spectrum}a.  

Bistability typically emerges only in the thermodynamic limit of infinite system size $L\to\infty$, see Fig. \ref{fig:damping-spectrum}b. An example in the context of driven Rydberg lattice gases is the formation of anti-ferromagnetically ordered steady states under conditions of Rydberg blockade \cite{Hoening2014}. Here a phase transition from
an unordered steady state into one with two different checkerboard-like distribution of Rydberg excitations occurs above a certain critical value of the optical 
driving associated with a spontaneous breaking of translational lattice symmetry.
For any finite system with linear dimension $L<\infty $ there is a unique steady state which is an equally weighted mixture of the two checkerboard configurations.
Each of these two configurations corresponds however to a metastable state, which, as shown in \cite{Macieszczak2016} are admixtures of the true steady state $\rho_0$ and the first excited eigenvector $\rho_1$. The metastable states
eventually relax to the true steady state on a time scale 
\begin{equation}
T \sim L^{\alpha>0}.
\end{equation}
Approaching the thermodynamic limit there is a critical slow down and the system becomes truly bistable. 

The transition from a unique to a bistable steady state is a true phase transition, which can be characterized by an order parameter.
Bistability and metastability cannot be distinguished, however, in finite systems and one has 
to investigate the scaling behavior of the characteristic timescales for $L\to \infty$. We will show now that bistability is always associated with long range order.
To this end we consider a translationally invariant system and assume that two orthogonal steady states, $\rho_0^{(1)}$ and $\rho_0^{(2)}$, exist. Then the asymptotic state at large times  is given by eq. \eqref{eq:bistable-ss}. In order to detect bistability there must be a local observable $\hat x_j$ that distinguishes the two states. Here the index $j$ labels positions (or some compact part of space) where $\hat x_j$ acts on.
Without loss of generality we assume
\begin{eqnarray*}
\langle \hat x_j\rangle _1 &=& {\rm Tr}\{\rho_0^{(1)} \hat x_j\} = +1,\\
\langle \hat x_j\rangle _2 &=& {\rm Tr}\{\rho_0^{(2)} \hat x_j\} = -1.
\end{eqnarray*}
Making use of translational invariance we find in the asymptotic state $\rho_{\rm ss}$, eq. \eqref{eq:bistable-ss}: $\langle \hat x_j\rangle_{\rm ss} = 2 p-1$.
Let us now consider $\hat Y = \sum_{j=1}^L  \left(\hat x_j-\langle \hat x_j\rangle\right)\sum_{k=1}^L\left(\hat x_k-\langle \hat x_k\rangle\right)$.
Obviously $\langle \hat Y\rangle \ge 0$ in any state. In both steady states, $\rho_0^{(\mu)}$ one finds
$\sum_{j,k} \langle \hat x_j\hat x_k\rangle_\mu \ge \sum_j \langle \hat x_j\rangle_\mu \sum_k \langle \hat x_k\rangle_\mu = L^2 $.
This yields for the general asymptotic state $\rho_{\rm ss}$
\begin{align}
\sum_{j,k} \langle \hat x_j\hat x_k\rangle_{\rm ss} &= p \sum_{j,k} \langle \hat x_j\hat x_k\rangle_{1} +(1-p)  \sum_{j,k} \langle \hat x_j\hat x_k\rangle_{2} \nonumber \\
&\ge L^2.
\end{align}
Now, considering correlations of the local observable $\langle\langle \hat x_j\hat x_k\rangle\rangle = \langle \hat x_j\hat x_k\rangle-\langle \hat x_j\rangle\langle\hat x_k\rangle$, which due to translational invariance only depend on $d=j-k$, one finds
\begin{equation}
\sum_{d} \langle\langle \hat x_0\hat x_d\rangle\rangle_{\rm ss} \ge L\left(1-(2p-1)^2\right).\nonumber
\end{equation}
Unless $p=0$ or $1$, this inequality can only be fulfilled in the thermodynamic limit $L\rightarrow \infty$ if
\begin{equation}
 \langle\langle \hat x_0\hat x_d\rangle\rangle_{\rm ss} \, \xrightarrow{d \rightarrow \infty}\, c \ne 0.
\end{equation}
Thus bistability is only possible if there is long-range order. This result is very intuitive. 
If only two (or some finite number) of stationary states exists, which differ in the expectation value of some local observable,
there must be long-range correlations in the system since one end of the system has to "know" in which state the other end is. For any system which is bistable in the thermodynamic limit we thus expect a scaling of the correlation length
\begin{equation}
\xi \sim L^{\beta>0}.
\end{equation}

Since the speed at which correlations
can spread in the system is finite, the transition into a bistable state is associated with a critical slow down, i.e. with
characteristic time scales that diverge with the system size.
If the correlation length $\xi$ is finite, the number of stationary configurations is $2^N$ with $N \sim \xi/L$. For $L\gg \xi$ the central
limit theorem holds and global probability distributions become single-peaked Gaussians. We will argue in the following that under
realistic experimental conditions correlation length and characteristic time scales in off-resonantly driven Rydberg lattice gases remain finite
and there is no bistability in this system.

\section{Experiment}
\label{sec:Experiment}

Experimentally, we continuously probe the Rydberg excitation dynamics in the anti-blockade regime. To this end, we prepare $\Nat = \num{20000}$ $^{87}$Rb atoms in a 3D optical lattice in the Mott insulating phase at unit filling. The lattice constants in $x$- and $y$- direction are $a_{x,y} = \SI{374}{\nm}$ and in $z$-direction $a_z = \SI{529}{\nm}$. We couple ground state atoms in the $5\mathrm{S}_{1/2}$ state continuously via a one-photon transition at a wavelength of $\lambda = \SI{297}{\nm}$ to the excited Rydberg state $25\mathrm{P}_{1/2}$ with Rabi frequency $\Omega$ and blue detuning $\Delta>0$. See Fig.\,\ref{fig:Singlerun}a for a sketch of the experiment. The spontaneous decay rate from the Rydberg state is given by $\Gs = \ts^{-1}=\SI{50}{\kHz}$ and we estimate a bare decoherence rate of $\gamma_0 \simeq \SI{300}{\kHz}$ which primarily originates from the laser linewidth.

The dipole trap lasers provide a weak photoionization rate $\Gamma_\mathrm{ion} \simeq \SI{2}{\kHz}$, which we use to observe the excitation dynamics in the system. The ions are guided towards a detector by a small electric field of $\SI{90}{\mV\per\cm}$. The detector efficiency is $40\%$. The retrieved ion signal with rate $R_\mathrm{ion}(t) = \Gamma_\mathrm{ion} \Nry(t)$, corrected by the detector efficiency, serves as a weak probe for the full excitation number $\Nry(t)$. We experimentally determine the Rabi frequency by analyzing the temporal statistics of the first detected ion. A more detailed discussion of the experimental setup and preparation can be found in \cite{Manthey2014} and in appendix \ref{ap:A}.  

In the inset of Fig.~\ref{fig:Singlerun}b the measured ion rate $R_\mathrm{ion}(t)$ for a Rabi frequency of $\Omega = 2\pi\times\SI{77}{\kHz}$ and a detuning of $\Delta = 2\pi\times\SI{13}{\MHz}$ is shown. The recorded signal extends over several tens of $\si{\ms}$, which is three orders of magnitude larger than the effective natural lifetime $\ts = \SI{20}{\us}$ (reduced by black-body transitions into neighbouring states) of the $25\mathrm{P}_{1/2}$ state. Therefore, in contrast to previous studies \cite{Schempp2014, Malossi2014}, our measurements allow us to study the steady-state of the system \cite{footnote_decay}. 

The experimentally accessible quantities, which we use to characterize the formation and the dynamics of Rydberg clusters, are the average ion rate and the time-dependent two particle correlation function
\begin{equation}
\langle R_\mathrm{ion}(t)\rangle\quad\textrm{and}\quad g^{(2)}(\tau) = \frac{\langle R_\mathrm{ion}(t+\tau) R_\mathrm{ion}(t)\rangle}{\langle R_\mathrm{ion}(t+\tau)\rangle\langle R_\mathrm{ion}(t)\rangle},
\label{equ:g2}
\end{equation}
respectively.
Already in a single experimental realization (see inset Fig. \ref{fig:Singlerun}b) strongly bunched excitations are visible, indicating strong excitation number fluctuations. To quantify them, we extract 
$g^{(2)}(\tau)$ of the measured ion signals, see Fig. \ref{fig:Singlerun}b. For typical parameters we find a value of $g^{(2)}(0) \gg 1$. Moreover, we find an exponential decay of the correlation signal with time scale $\tau^{(2)}$ much larger than the lifetime of the $25\mathrm{P}_{1/2}$ Rydberg state. Crucially, the strong excitation bunching reflects two different timescales arising in the anti-blockade regime: While a first \emph{seed} excitation is strongly suppressed (indications thereof can bee seen in the inset of Fig. \ref{fig:Singlerun}b as large intervals of zero signal in a single experimental run), the excitation rate for a second atom can be enhanced in the presence of the first \cite{Gaerttner2013, Lesanovsky2014a, Simonelli2016}. On top of that, further facilitated excitations lead to the correlated growth of Rydberg excitations which form a cluster. This leads to a large bunching amplitude $\gtwozero$ and a long correlation time $\tau^{(2)}$, which measures the lifetime of a single cluster $\tc$, i.e. $\tc = \tau^{(2)}$.  The ion rate $R_\mathrm{ion}$ and the cluster lifetime $\tc$ form the basis of our further analysis.

\section{Single Cluster Dynamics}
\label{sec:ClusterModel}

\begin{figure}
	\begin{center}
	\includegraphics[width=\columnwidth]{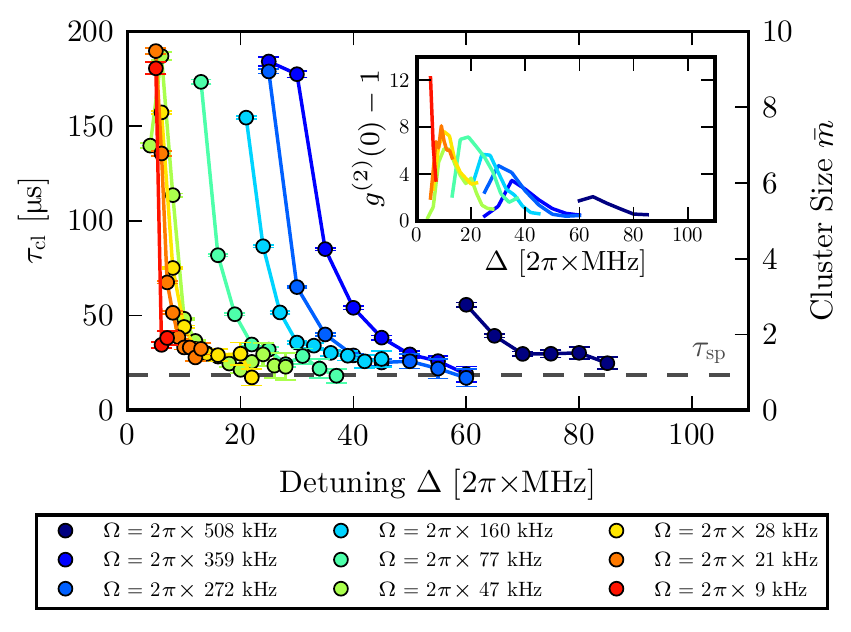}
	\end{center}
	\caption{Experimentally measured correlation time $\tau^{(2)}= \tc$ (left scale), corresponding to the lifetime of a cluster in the full anti-blockade regime $\Delta > 0$ for different Rabi frequencies $\Omega$. The cluster lifetime increases with decreasing detuning for fixed Rabi frequency. Using the cluster model we extract a typical cluster size $\bar m$ (right bar). The gray dashed line indicates the spontaneous decay time of a single atom $\ts = \SI{20}{\us}$. Inset: bunching amplitude $\gtwozero-1$ shows peaks at detunings which correspond to a cluster size of approximately two. Error bars correspond to the statistical uncertainty from fitting an exponential decay to the experimental retrieved correlation function. For simplicity they are not shown in the inset.
	} 
	\label{fig:ExperimentalResults}
\end{figure}
To gain insight into the cluster dynamics, we introduce a simplified cluster model and compare it to our experimental results. The model is capable to describe the growth of the excitation number in a single cluster and qualitatively captures all relevant features observed in the experiment. It characterizes the excitation cascade of independent, non-interacting excitation clusters, each of them being described by an effective rate equation model. 

In the case of strong decoherence it has been shown that rate equation models are a good approximation \cite{Ates2007a, Hoening2013, Petrosyan2013, Schoenleber2014} and were already successfully used to describe current experiments \cite{Schempp2014, Weber2015, Urvoy2015}. The effective cluster model is illustrated in Fig. \ref{fig:Singlerun}c. 
It only considers the total number of excitations $m$ and does not give insight into the internal microscopic structure of a cluster. The time evolution of the probability $p_m$ to be in a cluster of size $m$ is governed by the equations
\begin{align}
\partial_t p_0 =& -N_\mathrm{at}\Gamma_\mathrm{seed}p_0 + \Gamma_\mathrm{sp} p_1 \\
\partial_t p_1 =& +N_\mathrm{at}\Gamma_\mathrm{seed}p_0
- (\Gamma_\mathrm{sp}+\Gamma_\uparrow) p_1 \nonumber \\
&+ (\Gamma_\uparrow+2 \Gamma_\mathrm{sp} )p_2 \\
\partial_t p_m =& -(2 \Gamma_\uparrow + m\Gamma_\mathrm{sp}) p_m  + \Gamma_\uparrow p_{m-1} \nonumber \\
&+ \left(\Gamma_\uparrow + (m+1)\Gamma_\mathrm{sp} \right)p_{m+1} \quad (m\geq 2)
\end{align}

Within the single-cluster model we can identify three main contributions to the dynamics: 
\begin{enumerate}
  \renewcommand{\labelenumi}{(\roman{enumi})}
  \item a slow seed with rate $\Gamma_\mathrm{seed}$ for generating new clusters,
  \item a fast rate $\Gamma_\uparrow$ for increasing and decreasing the size of a cluster $m$ and
  \item the spontaneous decay rate $\Gs = \ts^{-1}$ limiting the size of a cluster.
\end{enumerate}
The seed excitation with rate 
\begin{equation}
\Gamma_\mathrm{seed} = \frac{2\Omega^2\gamma_0}{\gamma_0^2 + \Delta^2}
\label{eq:GammaSeed}
\end{equation}
is strongly suppressed due to the large detuning $\Delta$ in the anti-blockade regime. For a system with $\Nat$ atoms, we expect independent seed events to occur on a timescale $(\Nat\Gamma_\mathrm{seed})^{-1}$. However, the presence of already excited Rydberg atoms 
strongly alters the summed excitation rate of all neighboring atoms
\begin{equation}
\Gamma_\uparrow = \frac{ z_\mathrm{eff} 2\Omega^2\gamma_0}{ \gamma_0^2 + (\Delta-\Delta_\textrm{int})^2 },
\label{eq:GammaUp}
\end{equation}
where $\Delta_\textrm{int}$ is the interaction shift due to surrounding excited atoms. The competition between detuning $\Delta$ and interaction shift $\Delta_\mathrm{int}$ may lead to facilitated excitations \cite{Lesanovsky2014a}. The ratio between seed rate $\Gamma_\mathrm{seed}$ and the rate for cluster growth $\Gamma_\uparrow$, already allows us to identify different regimes of correlated excitation dynamics. For $\Gamma_\uparrow \gg \Gamma_\mathrm{seed}$, we expect a strong excitation cascade creating a cluster. In this regime each excitation triggers further excitations and we expect a substantial bunching amplitude $\gtwozero$. For $\Gamma_\uparrow \ll \Gamma_\mathrm{seed}$, we expect primarily uncorrelated excitations leading to a bunching amplitude $\gtwozero \simeq 1$ and clusters of size one. Finally we include spontaneous decay with rate $m \Gs$ increasing linearly with the cluster size $m$. Now, starting from an initial seed excitation, the cluster grows and shrinks with rate $\Gamma_\uparrow$ similar to a random walk along the cluster size $m \geq 0$.  Although, the single-atom spontaneous decay rate $\Gs$ can be small, this is the limiting factor for large Rydberg excitation clusters. Microscopically, a decay event creates a defect in a consecutive chain of excited atoms and splits the cluster into smaller parts. The strong interactions prevent that these parts can merge again. Besides the geometry we account for the splittings introducing an effective coordination number $z_\eff$. In appendix \ref{ap:B} we discuss the microscopic model in more detail.

The effective model allows us to compute quantitatively the typical size $\bar m$ of a cluster. Moreover, it gives us an intuitive understanding of $\gtwozero$ and permits us to estimate the number of independent clusters $\Nc$. Note, our further experimental analysis does not rely on the explicit evaluation of eqn. \eqref{eq:GammaUp} and is thus independent of the exact interaction potential between two Rydberg atoms.

\subsection{Cluster size}

To begin with, we discuss the typical size of a cluster $\bar{m}$. It is given by the ratio of the overall lifetime of a single cluster $\tc$ and the spontaneous lifetime of a single Rydberg excitation $\ts$, 
\begin{equation}
	\tc = \bar{m} \, \ts.
	\label{eq:tc}
\end{equation}
The above expression can be understood in the following way: When a cluster contains $\bar{m}$ excitations it requires $\bar m$ consecutive decay events of timescale $\ts$ before the whole cluster vanishes. The facilitation mechanism leads to an correlated excitation growth. This is in contrast to a non interacting system, where all excitations are independent and decay on the same single atom timescale $\ts$. We verified this intuitive picture by comparing full simulations to the simple cluster model and an analytic approach to the corresponding cluster size, see appendix \ref{ap:B}.
We use equation \eqref{eq:tc} to extract the cluster size from the experimental data. Within the single-cluster model, the cluster lifetime $\tc$ and thus the typical cluster size $\bar{m}$ only depend on the ratio $\Gamma_\uparrow/\Gamma_\mathrm{sp}$.

\begin{figure}
	\begin{center}
	\includegraphics[width=\columnwidth]{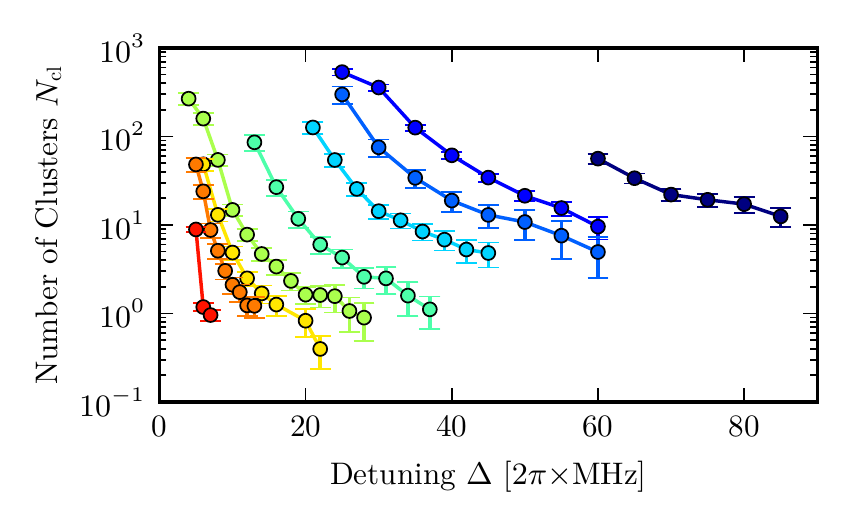}
	\end{center}
	\caption{Number of clusters $\Nc$ calculated from the experimental data using equation \eqref{equ:Clustersize} for the different detunings and Rabi frequencies. For decreasing detuning the number of clusters increases steadily, exceeding the size of a cluster by roughly one order of magnitude. The color code corresponds to the same Rabi frequencies as in Fig. \ref{fig:ExperimentalResults}. The errorbars correspond to the statistical uncertainties from fitting the cluster lifetime $\tc$ as well as the multi particle seed rate $\Nat\Gamma_\mathrm{seed}$.}
	\label{fig:Clusternumber}
\end{figure}

In Fig. \ref{fig:ExperimentalResults} the measured cluster size $\bar{m}$ (corresponding to a certain cluster lifetime $\tc$) is shown for the full spectrum in the anti-blockade regime ($\Delta > 0$) and different driving strength $\Omega$. The data reveal a strong increase in the cluster size by increasing the driving strength $\Omega$ as one would naively expect. In the limit of large detuning $\Delta \gg \Delta_\mathrm{int}$, all excitation rates are strongly suppressed and we obtain a cluster size of one corresponding to uncorrelated Rydberg excitations. In this case, the lifetime $\ts$ is the only relevant timescale. Close to the resonance, cascaded excitations lead to the formation of finite clusters of size $\bar{m} \lesssim 10$. Note that for strong driving and/or small detuning, the lifetime of the entire sample approaches the cluster lifetime, and no reliable values for $\tc$ can be extracted. The preferred generation of clusters with $\bar m > 1$ in this regime is associated with an excitation bunching. In the inset of Fig. \ref{fig:ExperimentalResults} we plot $\gtwozero-1$. Experimentally, we find substantial bunching amplitudes up to $\gtwozero \sim 13$. The measurements show a peak in $\gtwozero$ which coincides with a cluster size of $\bar{m} \simeq 2$. The peak shifts towards larger detuning with increasing Rabi frequency $\Omega$. The presence of a peak and its shift  can be understood in the following way: For larger detuning, where the typical cluster size is smaller than $\bar{m} = 2$, the influence of an uncorrelated background noise signal becomes important and diminishes the amplitude. For smaller detunings the cluster size as well as the rate of creating new independent clusters increases drastically. Both reduce the value of $\gtwozero$ due to an increase of uncorrelated ionization events.

\subsection{Number of clusters}

Next, we experimentally determine the seed rate $\Gamma_\mathrm{seed}$ by an analysis of the first ionization event in each single measurement (details can be found in appendix \ref{ap:A}). Comparing the timescale with which new clusters are produced $(\Nat\Gamma_\mathrm{seed})^{-1}$ to the lifetime of a single cluster $\tc$, we can deduce approximately the number of clusters $\Nc$ 
\begin{equation}
\Nc \simeq \Nat\Gamma_\mathrm{seed} \tc.
\label{equ:Clustersize}
\end{equation}
The results are shown in Fig \ref{fig:Clusternumber}. Similar to the cluster size $\bar{m}$, the number of clusters $\Nc$ increases with increasing driving strength $\Omega$ and decreasing detuning $\Delta$. This shows that for stronger driving and/or smaller detuning the Rydberg aggregate evolves into a steady state, which is characterized by the presence of a large fluctuating number of independent clusters with rather small size. This already suggests that in the experimentally observed parameter regime there is no global bistable phase with a large correlation length. For such a phase, we expect the system to be in a single cluster state with extent over the full system.

\subsection{Validity of single-cluster model}

\begin{figure}
	\begin{center}
	\includegraphics[width=\columnwidth]{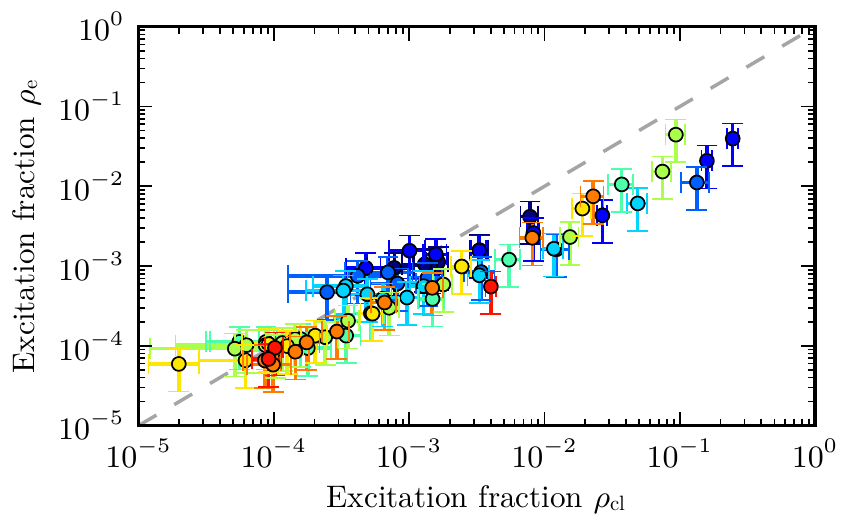}
	\end{center}
	\caption{Comparison between the mean excitation fraction $\rho_\mathrm{e}$ extracted from the maximal detected ion rate $R_\mathrm{ion}(t)$ and the excitation fraction $\rho_\mathrm{cl} = \Nc\bar{m}/\Nat$ calculated using the effective cluster model. The color code corresponds to the same Rabi frequencies as in Fig.~\ref{fig:ExperimentalResults}. The gray dashed line is a linear curve with a slope of one.}
	\label{fig:TheoryComparison}
\end{figure}

The extraction of the cluster size and number from the experimental data performed in the previous section
partially relies on the validity of the cluster model. In the following we make a consistency check to test the cluster model.
To this end, we compare the fraction of Rydberg excitations retrieved from the cluster model $\rho_\mathrm{cl} = \Nc\bar{m}/\Nat = \tc^2\Gs\Gamma_\mathrm{seed}$ to the excitation fraction $\rho_\mathrm{e}$ directly extracted from the measured ionization rate $R_\mathrm{ion}(t)$: $\rho_\mathrm{e} = R_\mathrm{ion}^\mathrm{max}/(\Nat\Gamma_\mathrm{ion})$. While the first expression incorporates the equations \eqref{eq:GammaSeed} and \eqref{eq:GammaUp} derived from the cluster model, the second solely relies on experimental data. Both quantities should be identical, independent on the Rabi frequency of the driving field and the detuning. The comparison is shown in Fig. \ref{fig:TheoryComparison}. The dashed line indicates the ideal agreement $\rho_\mathrm{cl} = \rho_\mathrm{e}$. 
We see that all experimental curves obtained for different driving strength fall on top of each other. For small excitation fractions, we find furthermore good agreement between the results from the effective cluster model and the experimental ionization signal. In the strong driving regime, the experimentally retrieved ion signal results in a smaller excitation fraction than predicted by the cluster model. This may have two reasons: Firstly, we monitor the ionization signal in a single experimental run with an exponential atom loss and take the maximal ionization signal to reflect the steady state excitation fraction without atom loss. Especially in the strong driving regime, atom loss may lead to smaller maximum ionization rates reducing the extracted excitation fraction $\rho_e < \rho_\mathrm{cl}$.
Secondly, our simple single-cluster model relies on the assumption that clusters are independent. However,  in appendix \ref{ap:B} we show by using a full rate equation simulation that cluster collisions become important for strong driving. Cluster collisions limit in particular the independent growth of clusters. This would also lead to a reduced excitation fraction $\rho_e < \rho_\mathrm{cl}$. In summary we conclude that the cluster model is an appropriate description for the cluster dynamics, especially for small excitation fractions.


\section{Beyond the Single Cluster Model}
\label{sec:Simulation}

The experimental protocol and the single cluster model introduced above only allow to measure and describe volume integrated quantities. To also study finite size effects,  we perform in this section numerical simulations using a many-body rate equation approach, which goes beyond the simple cluster model and can be extended to large system sizes. In particular we numerically perform a finite size extrapolation of the correlation time. In contrast to the cluster model, which describes "point-like" multi level systems, here we take into account the spatial distribution of the atoms in the experiment and calculate the level shift for each atom individually. In this way, we have access to the full spatial distribution of excitations and are able to track the growth and decay of every single cluster. The microscopic description and numerical details of the system are discussed in the following section \ref{sec:MircoscopicDetails}, while we here present the main results.

\subsection{Finite-size scaling of the correlation time}

\begin{figure}
	\begin{center}
	\includegraphics[width=\columnwidth]{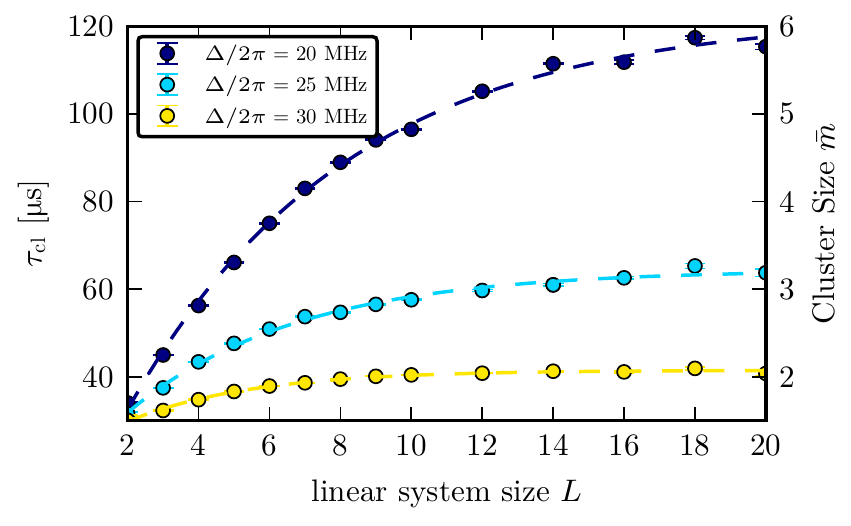}
	\end{center}
	\caption{Finite size extrapolation of the cluster lifetime $\tc$ for Rabi frequency $\Omega = 2\pi\times\SI{160}{\kHz}$ and different detuning $\Delta$. The dashed lines indicate an exponential saturation fit $\sim (1- c \exp(-L/L_0))$ with a constant $c$. Here, the characteristic length scale $L_0 = \num{6,2}, \num{3,2}$ and $\num{2,1}$ is comparable to the cluster size $\bar m$.} 
	\label{fig:SizeScaling}
\end{figure}

We use the rate equation model to study the finite size scaling of the cluster lifetime. The results are shown in Fig. \ref{fig:SizeScaling}. We increase the linear system size $L$ of a 3D Rydberg lattice gas and extract the cluster lifetime from the $\gtwotau$ correlation function for three different detunings and fixed Rabi frequency $\Omega = 2\pi \times \SI{160}{\kHz}$, using the same evaluation as for the experimental data. The data points strongly suggest a saturation in the limit $L \rightarrow \infty$. The dashed lines in Fig. \ref{fig:SizeScaling} correspond to an exponential saturation fit with a characteristic length scale matching the cluster size $\bar m$. This is already a first evidence that clusters, in general, do not extend over the whole system.

\subsection{Full counting distribution and hysteresis}

\begin{figure}
	\begin{center}
	\includegraphics[width=\columnwidth]{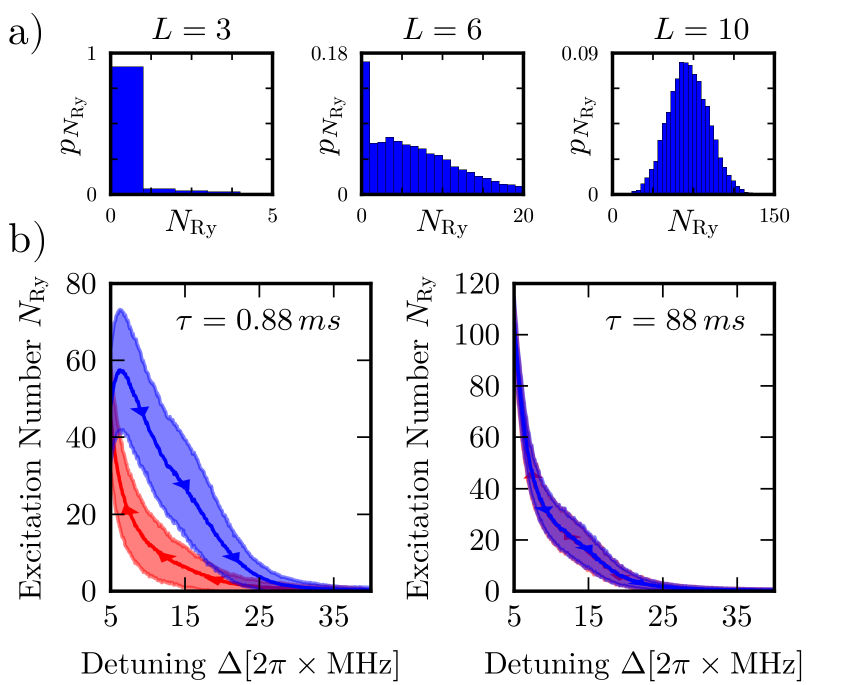}
	\end{center}
	\caption{ (a) Excitation number probability $p_{N_{\mathrm{Ry}}}$ for $\Omega = 2\pi\times\SI{185}{\kHz}$ and $\Delta = 2\pi\times\SI{15}{\MHz}$ for three different linear system sizes $L$ after \SI{1}{\ms} propagation. For intermediate system size ($L=6$) the distribution shows a bimodal behavior. (b) Hysteresis scan in the detuning parameter $\Delta$ for $\Omega = 2\pi \times \SI{160}{\kHz}$ with linear system size $L=10$. The sweep times in forward and backward direction are $\tau = \SI{0.88}{ms}$ and $\tau = \SI{88}{ms}$. In the case of a large sweep time $\tau = \SI{88}{ms}$ the dynamic hysteresis disappears. The mean Rydberg excitation number in the backward and forward scan are averaged over \num{500} trajectories. The standard deviation is indicated by the shaded area.} 
	\label{fig:SizeScaling-distribution}
\end{figure}

In a system where all timescales are finite, all correlation lengths must be finite unless the interactions are infinitely long-ranged. As a consequence
the bimodality of the full counting distribution resulting from the presence of two distinguishable metastable states should only persist if the system size
$L$ is smaller than the correlation length $\xi$. When $L$ exceeds $\xi$, averages are taken over independent regions and the bimodal distribution starts to wash out. In the limit $L\gg \xi$ the system behaves as 
$N=\left(L/\xi\right)^3$ independent sub-systems and the central limit theorem applies, eventually leading to an overall single-peaked Gaussian distribution. 
Fig. \ref{fig:SizeScaling-distribution}a shows the distribution function of the excitation number for different values of $L$ after $\SI{1}{\ms}$ excitation time. In a small system ($L = 3$), seed events are rare and the cluster size is limited by $L$. Therefore, the excitation number distribution is peaked at zero excitations. When the system size is comparable to the cluster extent ($L=6$), a bimodal excitation number distribution occurs. Here, a dynamical switching between low and high total excitation number is observable. However, in the limit of large systems ($L = 10$) the bimodal structure  disappears. Instead, a Gaussian distribution emerges with a peak at a large excitation number. Therefore, the bimodality seen in \cite{Malossi2014} may be a finite size effect. 
Our experimental results and theoretical investigations suggest that the correlation lengths remain short, on the order of the cluster extent and there is no phase transition to a global bistable regime.

Now, let us discuss dynamic hysteresis features present in metastable systems \cite{Casteels2016a}. While finite size effects vanish whenever the system size exceeds the correlation length $L \gg \xi$, see Fig. \ref{fig:SizeScaling-distribution}a, a hysteresis behavior can still be seen depending on the ratio between relaxation time $T$ and sweep time $\tau$. Here, the sweep time $\tau$ is the duration of a continuous parameter change in the detuning $\Delta$ from an initial to a final value. In our system, the relaxation towards the stationary state is determined by the growth of finite clusters. Therefore, we expect and numerically verified that $T$ is on the order of the cluster lifetime $\tc$. Exemplary, we show in Fig. \ref{fig:SizeScaling-distribution}b for two different sweep times $\tau$ the dynamic hysteresis. In the case of a small sweep time $\tau = \SI{0.88}{ms}$ a large dynamic hysteresis area can be identified, while for $\tau = \SI{88}{ms}$ the hysteresis area vanishes. The hysteresis behavior persists on a rather large timescale compared to the relaxation time, which is consistent with a simple model that describes the sweep as a sequence of successive small parameter quenches. Since the steady state excitation number $\bar{N}_\mathrm{Ry}(\Delta)$ is different for different values of $\Delta$, the coarse grained relaxation of the excitation number is no longer exponential, but attains algebraic corrections. In the limit of many small parameter quenches with approximately constant relaxation time $T$ starting from an initial detuning $\Delta_0$ to $\Delta_1$ in time $\tau$, the Rydberg excitation number is given by
\begin{equation}
\label{eq:HystereseRelaxation}
N_\mathrm{Ry}(\tau) = N_\mathrm{Ry}(0) e^{-\tau/T} + \tau/T \int_0^1 \mathrm{d}x \ \bar{N}_\mathrm{Ry}(x) e^{-\tau/T x},
\end{equation}
with $x = (\Delta_1-\Delta)/(\Delta_1-\Delta_0)$. For $\bar{N}_\mathrm{Ry} = \mathrm{const}$ we obtain the known exponential relaxation with timescale $T$. However, as seen in Fig. \ref{fig:SizeScaling-distribution}b, the mean Rydberg excitation number is typically not constant. Linearization of $\bar{N}_\mathrm{Ry}$ directly results in algebraic correction scaling with $(\tau/T)^{-1}$. This may have a tremendous effect on the hysterese relaxation. To give an example, in our system, the relaxation time is on the order of $T \sim \SI{100}{\us}$ comparable to the cluster lifetime. However, the hysterese area vanishes only on a timescale of $\sim \SI{100}{\ms}$, which is three orders of magnitude larger. The algebraic relaxation of the hysteresis area is consistent with other simulations \cite{Casteels2016a, Sibalic2016} and agrees well with the findings in our system. For more details on the hysterese relaxation see appendix \ref{ap:E}.

\section{Microscopic Description \& Simulation}
\label{sec:MircoscopicDetails}

Before we conclude, we will give some more details about the microscopic description of our experiment. For the relevant length scales we use an approximate interaction potential $V(r) \simeq C_9/r^9$ with $C_9 = 2\pi \times \SI{2.1}{\kHz \um^9}$, which is motivated by exact diagonalization of a subspace of the two body interaction Hamiltonian including dipole-dipole, dipole-quadrupole and quadrupole-quadrupole interactions \cite{Flannery2005}. For details regarding the calculation see appendix \ref{ap:inter}. We do not take into account possible black-body induced transitions into neighbouring Rydberg states, which have recently been suggested to explain anomalous line broadening in dense samples \cite{Goldschmidt2016} and which would introduce different types of interactions. While this effect probably is present on the timescales our experiments are performed on, we believe that because we continuously pump atoms into the 25P-state, the above given potential is still dominating the cluster growth dynamics. We did also check that the qualitative behavior of the system discussed here stays the same for a different type of interaction (e.g. for van der Waals-type potentials). 

Besides the decoherence stemming mainly from laser noise $\gamma_0$, we also include an inhomogeneous broadening $\gamma_\mathrm{inh}$ motivated by the coupling to motional states in a lattice of finite width \cite{Li2013, Macri2014}. Specifically, we use an additional approximate decoherence rate 
\begin{equation}
\gamma_\mathrm{inh} = \sum_{\langle j \rangle} |\partial_r V(r_j)| \sigma/\sqrt{\pi},
\end{equation}
 where $\sigma \simeq \SI{60}{\nm}$ is the width of the localized wavepacket in the optical lattice and the sum runs over all excited atoms $\langle j \rangle$ with distance $r_j$. Therefore, the total decoherence rate is $\gamma = \gamma_0 + \gamma_\mathrm{inh}$. For our interaction potential this is a good approximation for $\Delta \gtrsim 2\pi\times\SI{18}{\MHz}$. (For a detailed discussion see appendix \ref{ap:C}.) Additionally, we take into account the ionization process by a rate $\Gamma_\mathrm{ion}$, which corresponds to a loss of atoms. Each ionization event alters the geometry of the sample of Rydberg atoms. To properly account for the detection background, we superimpose the extracted Rydberg excitation dynamics with an uncorrelated noise signal with rate $\Gamma_\mathrm{n}$. It originates from atoms which are trapped on the outer region of the optical lattice. Since the atomic density drops exponentially, these atoms do not contribute to the cascaded excitation dynamics. Although, the background noise rate $\Gamma_\mathrm{n}$ may depend on the detuning $\Delta$ and driving strength $\Omega$, it is sufficient to approximate it with a constant value $\Gamma_\mathrm{n} \simeq \SI{1}{\kHz}$ here. Note that all parameters used in our simulations are estimated from or determined by our experimental measurements.  For a typical simulation using $\Nat = \num{1000}$ atoms we already see strongly reduced finite size effects.

\begin{figure}
	\begin{center}
	\includegraphics[width=\columnwidth]{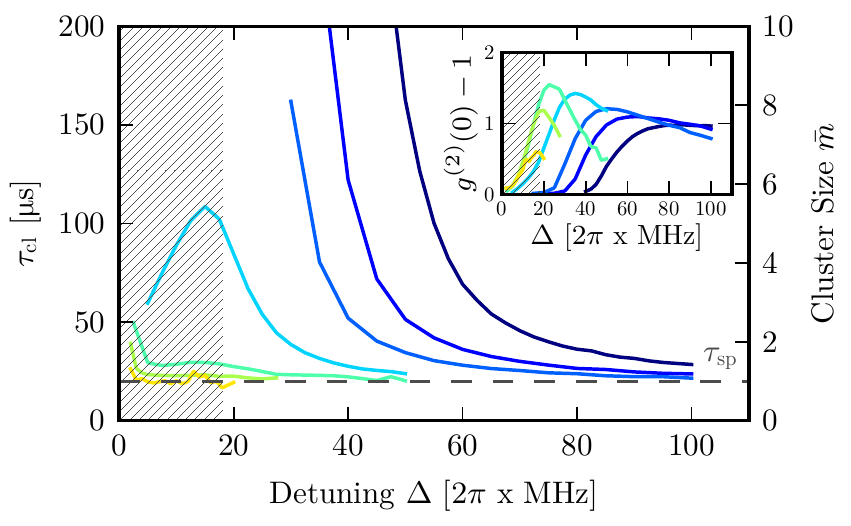}
	\end{center}
	\caption{Numerical results of the cluster lifetime $\tc$ and the bunching amplitude $\gtwozero-1$ (inset) using a full many-body rate equation model with parameters estimated from the experiment. The color code corresponds to the same Rabi frequencies as in Fig.~\ref{fig:ExperimentalResults}. The system contains $\Nat = 1000$ atoms. We use an interaction potential $V(r) = C_9/r^9$ with $C_9 = 2\pi\times\SI{2,1}{\kHz\um^9}$. Other parameters are: $\Gamma_\mathrm{ion} = \SI{1}{\kHz}$, $\gamma_{0} = \SI{300}{\kHz}$, $\Gamma_\mathrm{n} = \SI{1}{\kHz}$ and $\sigma = \SI{0,06}{\um}$. The gray shaded area corresponds to the area where the decoherence rate $\gamma_\mathrm{inh}$ is not a good approximation anymore (see appendix \ref{ap:C}).} 
	\label{fig:fullRateResults}
\end{figure}

The numerical results obtained for the temporal correlation function $\gtwotau$ are shown in Fig. \ref{fig:fullRateResults}. Although, the simulations do not reproduce the experimental results on a quantitative level, we find good qualitative agreement. Firstly, the typical cluster sizes are consistent with the values obtained in the experiment and have approximately the same behavior when varying $\Delta$ and $\Omega$. Secondly, the structure of the numerically evaluated bunching amplitude $\gtwozero$ is similar to the one extracted from experiment. In particular, the $\gtwozero$ peak coincides with a cluster size of $\bar{m} = 2$ as seen experimentally. However, the actual size of $\gtwozero$ depends on the system size and therefore strongly deviates from the experiment. We believe this originates from e.g. the motion of the excited atoms due to the interaction or the decay to other Rydberg states due to black-body radiation. While we are aware that these effects could play a role it is numerically very challenging to incorporate all of them in a full many-body simulation. From the results obtained we do infer that the rate equation model used is nevertheless a sufficiently good approximation of the Rydberg excitation dynamics seen in the experiment.

After checking the validity of the numerical simulation, we can employ it to theoretically study the spatial cluster dynamics. In the case of a van-der-Waals interaction $V(r) = C_6/r^6$ we observe directional excitation dynamics as indicated in Fig. \ref{fig:Singlerun}a along a certain lattice direction. The directionality stems from the competition between the quasi long range interaction and the finite linewidth $\gamma$. To understand this, let us discuss the case of two neighboring Rydberg excitations on a lattice where a single excitation facilitates its nearest neighbors, i.e. $\Delta = V(a)$. Now, for a third particle along the direction of the two excitations, the residual interaction shift in anti-blockade configuration is $V(a)/64$, while in the orthogonal direction it is $V(a)/8$. Comparing to the linewidth $\gamma$, this may suppress the growth of the cluster in orthogonal directions while promoting the growth along the lattice direction. Crucially, the directionality depends on the range of the interaction. In the case of short range interaction, e.g. the interaction potential $V(r) = C_9/r^9$ relevant for our experiment, the impact on the next nearest neighbors are typically negligible. Therefore, no directional Rydberg excitation dynamics is seen in the full numerical simulations using this potential. However, we believe that directional growth should be experimentally observable in different systems.

\section{Conclusion \& Outlook}

To conclude, we discussed the excitation dynamics of a large dissipative Rydberg lattice gas probing the full anti-blockade regime. We showed that the increase in the relaxation time corresponds to the formation of small clusters within the Rydberg aggregate. Using a simplified cluster model, we could estimate the size of individual clusters $\bar m \lesssim 10$ as well as the number of independent clusters $\Nc \lesssim 500$. Furthermore, we experimentally tested the validity of the effective cluster model. Using a many-body rate equation model, we find qualitative agreement between our theoretical model and the experimental results. 
We performed an extrapolation of the cluster size with the linear system size indicating that the clusters remain small. The absence of long-range correlations indicates the absence of a true phase transition to a bistable regime. Nevertheless, a bimodal excitation number distribution can be observed for small system sizes, as well as a dynamic hysteresis for small sweep times.

Our results show that the experimental and theoretical identification of phase transitions in open quantum systems require a careful analysis of the system size scaling and of the resulting correlation functions. To fully understand the thermodynamic limit of such systems, numerical simulations should be benchmarked with experimental data. Given the huge numerical effort to calculate and the experimental challenge to measure higher order correlation functions with good spatial and temporal resolution, this is in general a non-trivial task. 
We believe that temporal correlations are very helpful in this respect as they directly reflect the intrinsic dynamics and fluctuations of the steady state. Spatial correlations instead are typically limited to single destructive snapshots of the system and are insensitive to the dynamics.

In our system a transition to a truly bistable stationary phase is expected to be of first order \cite{Ates2012}. An experimental realization would allow to study
these types of phase transitions in stationary states of open systems. We have shown that anti-blockaded Rydberg gases with van-der Waals type interactions 
do not lead to diverging correlation times and long-range order in the experimentally relevant regime of large decoherence. Thus there is no phase transition  
to a bistable phase. Whether or not long range order can occur in a coherent regime and taking into account atomic motion remains an open question. Numerical simulations in this regime 
can be performed only for very small system sizes as the coherent many-body dynamics is no longer accessible by classical Monte-Carlo simulations. Here
experiments are needed, which however require a substantially reduced decoherence level.

\section*{Acknowledgments}
We acknowledge financial support by the DFG within the SFB/TR185. We thank Carsten Lippe for discussing and revising the manuscript. We thank Dominik Linzner and David Petrosyan for stimulating discussions. F.L. and O.T. are recipients of a fellowship through the Excellence Initiative MAINZ (DFG/GSC 266). This work was supported by the Allianz für Hochleistungsrechnen Rheinland-Pfalz (AHRP).

\section*{Author Contributions}
M.F. and H.O. conceived the study of off resonant Rydberg excitation dynamics. O.T and T.N. performed the experiment and analyzed the data. H.O. supervised the experiment. F.L. and M.F. developed the theoretical model. F.L. performed the numerical simulations of the many-body rate equation model. F.L. and O.T. performed numerical simulations of the single cluster model. M.F. supervised the numerical simulations. F.L. and O.T. prepared the early version of the manuscript. All authors contributed to the data interpretation and manuscript.

\appendix

\section{Experimental Methods}
\label{ap:A}
An experimental cycle consists of loading a 3D magneto optical trap of $^{87}$Rb for \SI{2,5}{\s} from a 2D magneto optical trap. The precooled atoms are loaded into a crossed optical dipole trap, generated from a Nd:YAG fiber amplifier at \SI{1064}{\nm}, and undergo forced evaporation within \SI{4}{\s}, resulting in a Bose-Einstein condensate of \num{20000} atoms in an isotropic trap with a trap frequency of $2\pi\times\SI{64}{\Hz}$. After evaporation we perform a \SI{2}{\ms} long linear scan with a scanning electron microscope across the cloud, yielding an ion signal proportional to the atom number while only marginally influencing the atomic sample. This allows us to check for atom number fluctuations during preparation and to correct for long term instabilities. Because of a finite magnetic field gradient present during evaporation, the sample is spin polarized in the $F=1, m_F = +1$ state from which we transfer it to the $F=2, m_F = +2$ state by a microwave Landau-Zener sweep. We load the prepared condensate into a 3D optical lattice, created from a Ti:sapphire laser at \SI{748}{\nm}, with lattice constants $a_{x,y} = \SI{374}{\nm}$ and $a_z = \SI{529}{\nm}$ in an exponential ramp with time constant $\tau = \SI{20}{\ms}$ to a final lattice depth of $S = \num{20}E_\mathrm{rec}$. This ensures that we are in the Mott insulating phase regime and have a maximum of one atom per lattice site \cite{Manthey2015}.

To measure the cluster dynamics of Rydberg aggregates we couple the atomic cloud with a one-photon transition at \SI{297}{\nm} to the Rydberg state $25\mathrm{P}_{1/2}$ with fixed detuning $\Delta$ and intensity $I$ for \SI{100}{\ms}. The laser light is produced from a frequency doubled dye laser at $\SI{594}{\nm}$. During the light-matter interaction, Rydberg excitations generated in the cloud are photoionized by the trapping light and black-body radiation with a rate of $\Gamma_\mathrm{ion} \simeq \SI{2}{\kHz}$ and guided by a small electric field of \SI{90}{\mV\per\cm} to a discrete dynode detector. The underlying lattice structure ensures hereby that effects from associative ionization \cite{Niederprum2015} or molecular Rydberg states \cite{Manthey2015} are negligible. Nevertheless, we only probe with blue detuning where molecular states are only addressable through a spin-flip in the second atom \cite{Niederpruem2016}, which is strongly suppressed compared to a normal molecular excitation. We analyse the detected ion signal by calculating the $\gtwo$-correlation function binned with \SI{1}{\us}. We correct for the overall decay of the sample by normalizing with the averaged ion rate of multiple runs. Because of detector ringing and ion repulsion during the time of flight to the detector we can't analyse the first two microseconds of the correlation function.

\begin{figure}
	\begin{center}
	\includegraphics[width=\columnwidth]{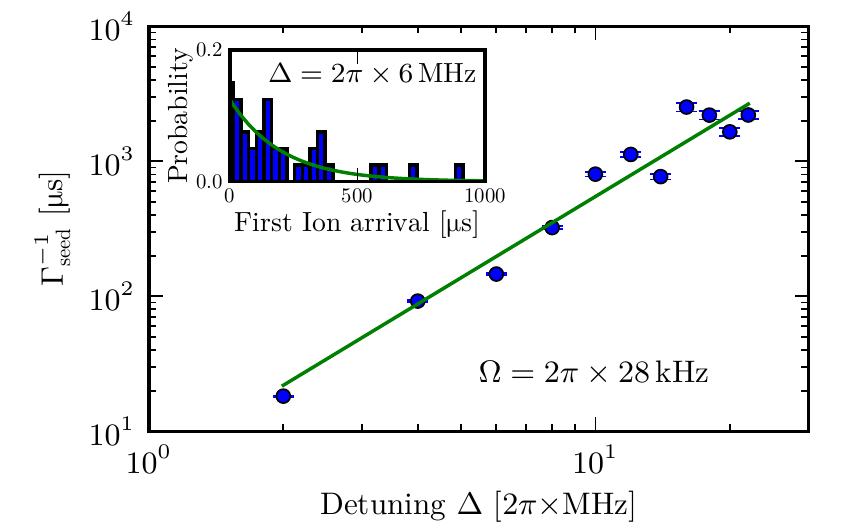}
	\end{center}
	\caption{We retrieve the seed rate $\Gamma_\mathrm{seed}$ by fitting an exponential decay $\exp(-\Gamma_\mathrm{seed}t)$ to the probability of detecting a first ion after the time of flight $t$ to our ion detector (inset). From a $\sim \Delta^2$ fit to the extracted seed rates we can calculate the Rabi frequency $\Omega$, taking into account the finite detection and ionization efficiency, as well as the bare decoherence rate.}
	\label{fig:apA-Fig1}
\end{figure}

The seed rate $\Gamma_\mathrm{seed}$ and Rabi frequency $\Omega$ for each experimental parameter are determined by an analysis of the arrival time of the first detected ion (see Fig. \ref{fig:apA-Fig1}). Note that the first ion analysis is independent from the cluster dynamics analysis. By doing a statistical analysis for each laser power and detuning over multiple experimental runs we can extract the multi-particle seed rate, taking into account our finite detection and ionization efficiency. Additionally, we extract the underlying Rabi frequency by fitting the seed rates for different detuning with $\Gamma_\mathrm{seed} = \frac{2\Omega^2\gamma}{\Delta^2}$ (for $\Delta \gg \gamma$). We checked that the extracted Rabi frequencies approximately follow the expected $\sqrt{I}$ scaling.

\section{Cluster Model}
\label{ap:B}

\begin{figure*}
\centering
\epsfig{file=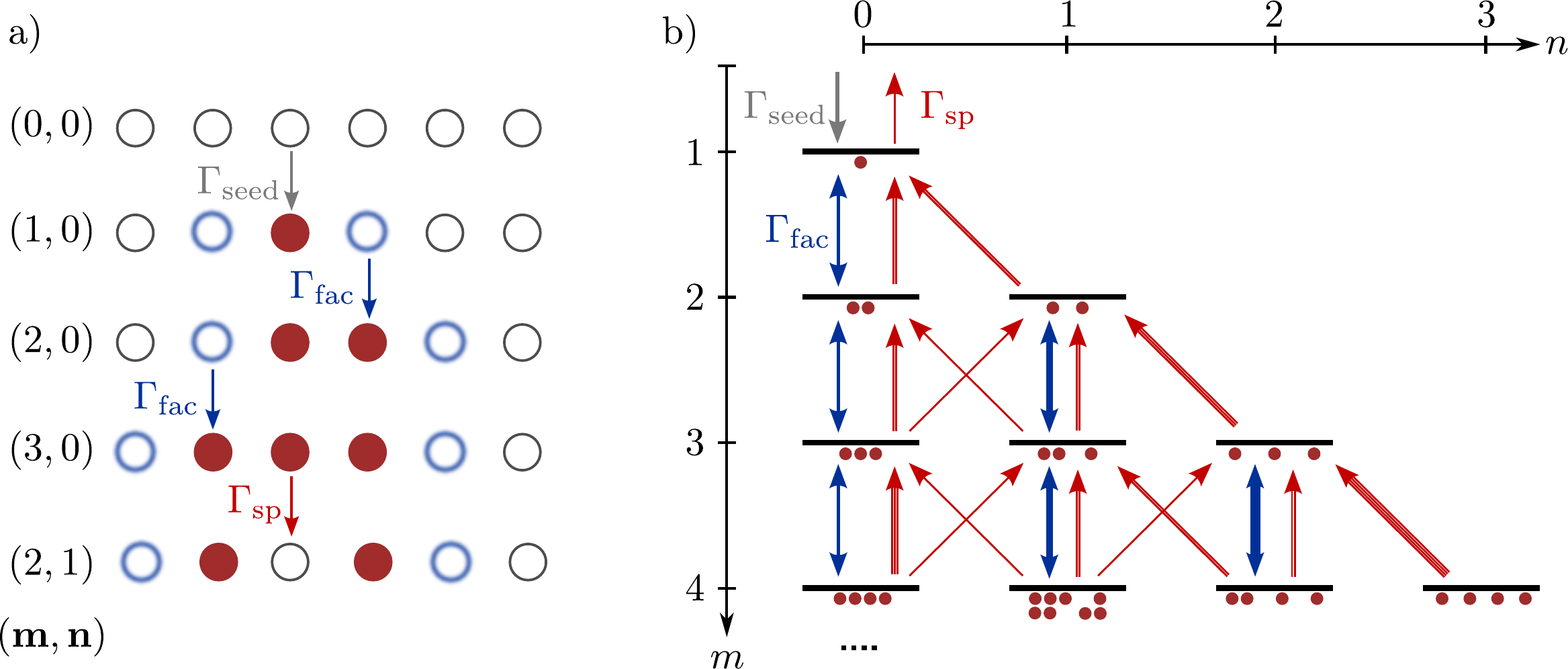, width=.9\textwidth}
\caption{(a) Different configurations during the time evolution of a single cluster starting from an initial seed excitation. In the last configuration, the two excited atoms prevent the excitation of the atom in the middle. A cluster, which is split apart, can not merge. (b) The configuration space of the cluster dynamics is spanned by the cluster size $m$ and the number of splittings $n$. }
\label{fig:apB-Fig1}
\end{figure*}

Here, we study a simple 1D lattice of Rydberg atoms in anti-blockade configuration. Specifically, we set the lattice constant $a$ equal to the facilitation radius $r_\mathrm{fac} = \left( C_6/\Delta \right)^{1/6}$ assuming that Rydberg atoms interact via a typical van der Waals potential $V(r) = C_6/r^6$. We compare a full numerical simulation based on a many-body rate equation model \cite{Ates2007a, Hoening2013, Petrosyan2013, Schoenleber2014} to our simplified cluster model introduced in Sec. \ref{sec:ClusterModel}.

\begin{figure}
\centering
\epsfig{file=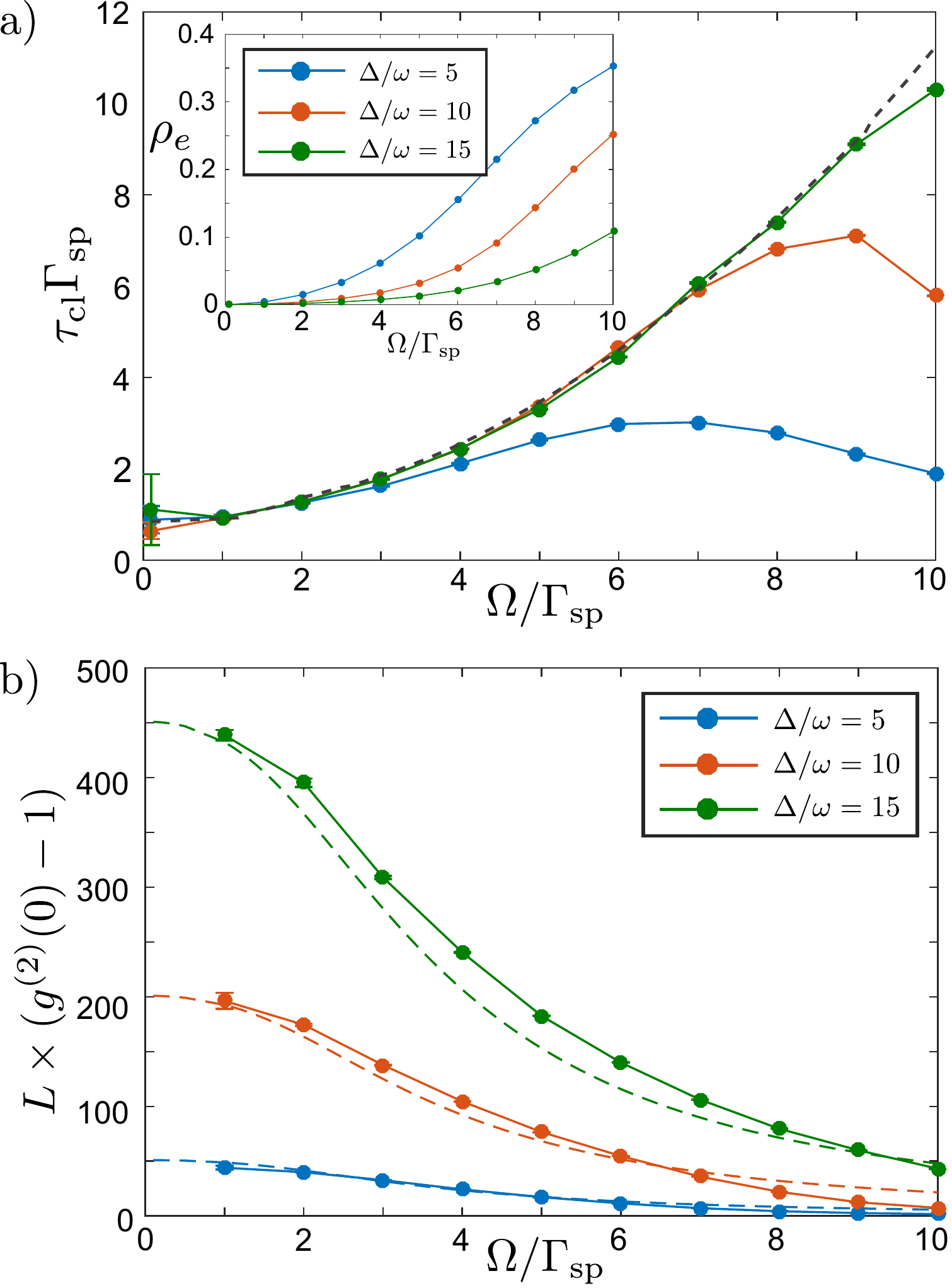, width=.42\textwidth}
\caption{ Numerical simulation of the dynamics using a many-body rate equation method in a large 1D lattice system with size $L \gg 100$ for different detunings $\Delta$ over the linewidth $\omega = \gamma_0\sqrt{\Omega^2/\gamma_0\Gs+1}$ and fixed facilitation radius $r_\mathrm{fac} = a$. The decoherence rate is $\gamma_0/\Omega = 2.05$ and decay rate is $\Gs/\Omega = 0.1$. (a) Cluster lifetime $\tc$ and (b) bunching parameter $g^{(2)}(0)-1$. The dashed lines are the results from the cluster model using $z_\eff \simeq 2+(\Omega/\Gs)^2/8$ as the effective coordination number.}
\label{fig:apB-Fig2}
\end{figure}

As pointed out before, the first seed excitation is produced with a rate $\Nat\Gamma_\mathrm{seed}$. Now, this seed excitation triggers further excitations with a rate proportional to $\Gamma_\mathrm{fac} \gg \Gamma_\mathrm{seed}$, where
\begin{equation}
\label{eq:FacilitationRate}
\Gamma_\mathrm{fac} = \frac{2\Omega^2}{\gamma_0}.
\end{equation}
In the case of a 1D system, the first excitation enhances the excitation rate of its two neighbors, see Fig. \ref{fig:apB-Fig1}a. We introduce a geometric coordination number $z$ counting the number of atoms with enhanced excitation rate \eqref{eq:FacilitationRate}. In the 1D case we identify a geometric coordination number $z = 2$ for the transition between cluster size $m=1$ to $m=2$, see Fig. \ref{fig:apB-Fig1}b. First, let us neglect the effect of long-range interactions beyond the lattice constant $a$. Then, a cluster grows and shrinks with rate $z\Gamma_\mathrm{fac}$. This is in analogy to a random walk along the cluster size axis $m$ with spreading $\propto \sqrt{\Gamma_\mathrm{fac}t}$. However, due to spontaneous decay with rate $\Gs$, the size of a cluster $m$ is limited. The competition between drive and decay leads to a finite size. A decay event may split the cluster into two parts, see Fig. \ref{fig:apB-Fig1}b, leading to a doubling of the coordination number $z$. However, this requires that the individual parts of the cluster are spatially well separated and do not influence each other. For instance, the last configuration in Fig. \ref{fig:apB-Fig1}a shows that the two parts of the cluster can not merge into a single one. Without the effect of two atoms blocking each other, the coordination number $z \simeq 2n+1$ increases along the axis of splittings $n$. In Fig. \ref{fig:apB-Fig1}b, the transition rates between all possible configurations labeled by cluster size $m$ and number of splittings $n$ are shown.

In the following, we discuss assumptions, which allow us to estimate the size and the lifetime of a cluster. Since we are mostly interested in the cluster size $m$, we firstly neglect all splitting processes fixing the dynamics to the case $n=0$. This is correct in the regime of weak driving $\Gamma_\mathrm{fac} \simeq \Gs$ with coordination number $z=2$ due to the geometry. For strong driving $\Gamma_\mathrm{fac} \geq \Gs$, we have to account for an increase in the coordination number $z$. To do so, we determine an effective coordination number $z_\eff$ self-consistently. This results in the simplified cluster model discussed in Sec. \ref{sec:ClusterModel} with $\Gamma_\uparrow = z_\mathrm{eff}\Gamma_\mathrm{fac}$, see Fig. \ref{fig:Singlerun}c. Now, using a detailed balance ansatz and projecting to the case of having a cluster ($m \geq 1$), we calculate the cluster size distribution $p_m$. This yields the iterative formula
\begin{equation}
p_m = \frac{z_\eff\Gamma_\mathrm{fac}}{z_\eff\Gamma_\mathrm{fac}+m\Gs} p_{m-1}
\end{equation}
with normalization condition $\sum_{m=1} p_m = 1$. Using the full width half maximum value $\bar m$ of the cluster size distribution $p_m$, we can estimate the cluster lifetime $\tc$ to be equal with equation \eqref{eq:tc}:
\begin{equation}
\tc \simeq \bar m / \Gs.
\end{equation}
Note that the cluster lifetime agrees with the lifetime of a single Rydberg excitation in the limit $\Gamma_\mathrm{fac} \ll \Gs$, i.e. $\bar m = 1$. 

Now, we compare the dynamical properties of a large lattice system using a many-body rate equation method to the simplified cluster model. As discussed in Sec. \ref{sec:Experiment}, to obtain the lifetime of a single cluster, we calculate the second order temporal correlation function $g^{(2)}(\tau)$. We extract the relaxation time, which we identify with the cluster lifetime, and the bunching parameter $g^{(2)}(0)$. The results for different detunings $\Delta$ and fixed facilitation radius $r_\mathrm{fac} = a$ are shown in Fig. \ref{fig:apB-Fig2}. As we would expect, with increasing driving strength $\Omega/\Gs$ the lifetime of the cluster increases. This allows us to determine the effective coordination number $z_\eff$ used in the simplified cluster model. Besides the geometric coordination number $z = 2$, we account for the number of splittings, which occur with increasing cluster lifetime, by setting 
\begin{equation}
z_\eff \simeq z + (\tc - \ts)/\ts,
\end{equation}
where $\ts=\Gs^{-1}$ is the lifetime of a single Rydberg excitation. Comparing the effective cluster model (dashed line) with the full rate equation simulations (bullet points), we find excellent agreement for small driving strength $\Omega/\Gs$. However, the full simulations show that the cluster lifetime decreases again 
with $\Omega/\Gs$
depending on the detuning $\Delta$. We can understand this as an effect of interacting clusters in the high excitation density regime, which is beyond the simple cluster model. Increasing the detuning $\Delta$, we decrease the seed rate $\Gamma_\mathrm{seed}$ with which new clusters are produced. Therefore, the Rydberg excitation density $\rho_\mathrm{e}$ decreases, see inset Fig. \ref{fig:apB-Fig2}a. In our numerical simulation, we identify the regime where cluster collisions are important near an excitation density $\rho_\mathrm{e} \simeq 0.2$. These collisions strongly reduce the independent growth of a cluster.

Moreover, using the effective cluster model and including the first seed excitation in the cluster size distribution, we can estimate the bunching parameter $L \times (g^{(2)}(0)-1) = \langle m(m-1) \rangle/\langle m \rangle^2$. Comparing both, the full simulations with the simplified cluster model, we find good agreement. With increasing driving strength $\Omega/\Gs$, the bunching parameter, which measures deviations from Poisson counting statistics, decreases. Therefore, with growing $\Omega/\Gs$, the number of independent clusters $\Nc$ and the lifetime $\tc$ increases. Both contribute to a reduction of the bunching parameter.

\section{Rydberg-Rydberg interaction potential}
\label{ap:inter}

\begin{figure}
	\begin{center}
	\includegraphics[width=\columnwidth]{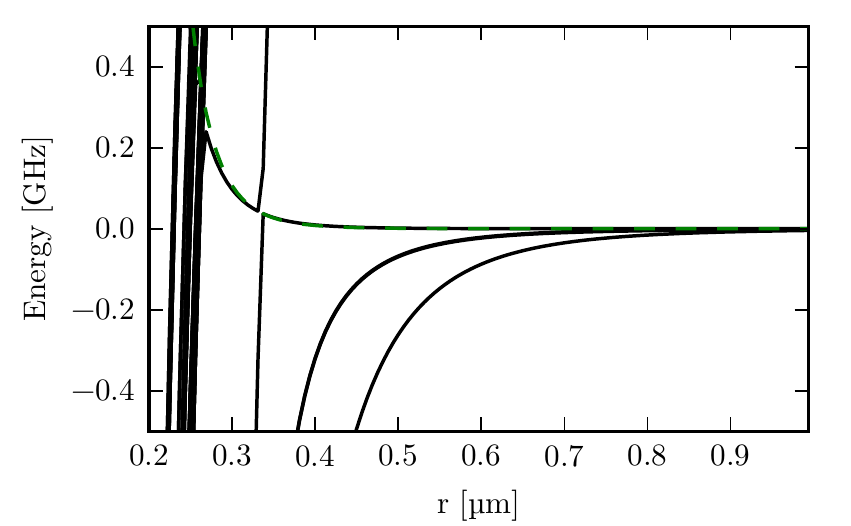}
	\end{center}
	\caption{Interaction potentials of the pair state $\ket{25\mathrm{P}_{1/2},25\mathrm{P}_{1/2}}$ obtained from a diagonalization of the interaction Hamiltonian including up to quadrupole-quadrupole interactions. The different potentials correspond to different superpositions of the $m_j$-states. The green dashed line is a fit to the only repulsive potential with $V(r) = C_9/r^9$ and $C_9 = 2\pi\times\SI{2.1}{\kHz\um^9}$.}
	\label{fig:apInter-Fig1}
\end{figure}

To calculate the interaction potential between two Rydberg atoms we perform an exact diagonalization of the interaction Hamiltonian as shown in \cite{Flannery2005}. We use a basis of pair states $\ket{n_1l_1j_1m_{j1}n_2l_2j_2m_{j2}}$ in the energetic vicinity of the state of interest and include all interaction terms up to quadrupole-quadrupole interactions as well as a small finite magnetic field of \SI{1}{Gs}. For the calculations shown here the basis consists of 1604 pair states with a maximum angular momentum quantum number of $l=3$ in an energetic vicinity of \SI{50}{\GHz} to the pair state $\ket{25\mathrm{P}_{1/2},25\mathrm{P}_{1/2}}$. The results of the diagonalization are shown in Fig. \ref{fig:apInter-Fig1}. For the $\ket{25\mathrm{P}_{1/2},25\mathrm{P}_{1/2}}$ state there are four different interaction channels depending on the $m_j$ state of the pair state. The symmetric superposition $\frac{1}{\sqrt{2}}\left(\ket{+1/2,-1/2}+\ket{-1/2,+1/2}\right)$ shows the strongest attraction, the two fully stretched states are nearly degenerate but still attractive, while the antisymmetric superposition gets blue shifted by an dipole-quadrupole interaction with the $\ket{22\mathrm{F}_{5/2}24\mathrm{D}_{3/2}}$ $m_j$-manifold. We fit the relevant repulsive interaction potential with a $C_9/r^9$-type potential and use it in the simulations presented here.

\section{Inhomogeneous Broadening}
\label{ap:C}

\begin{figure}
\centering
\epsfig{file=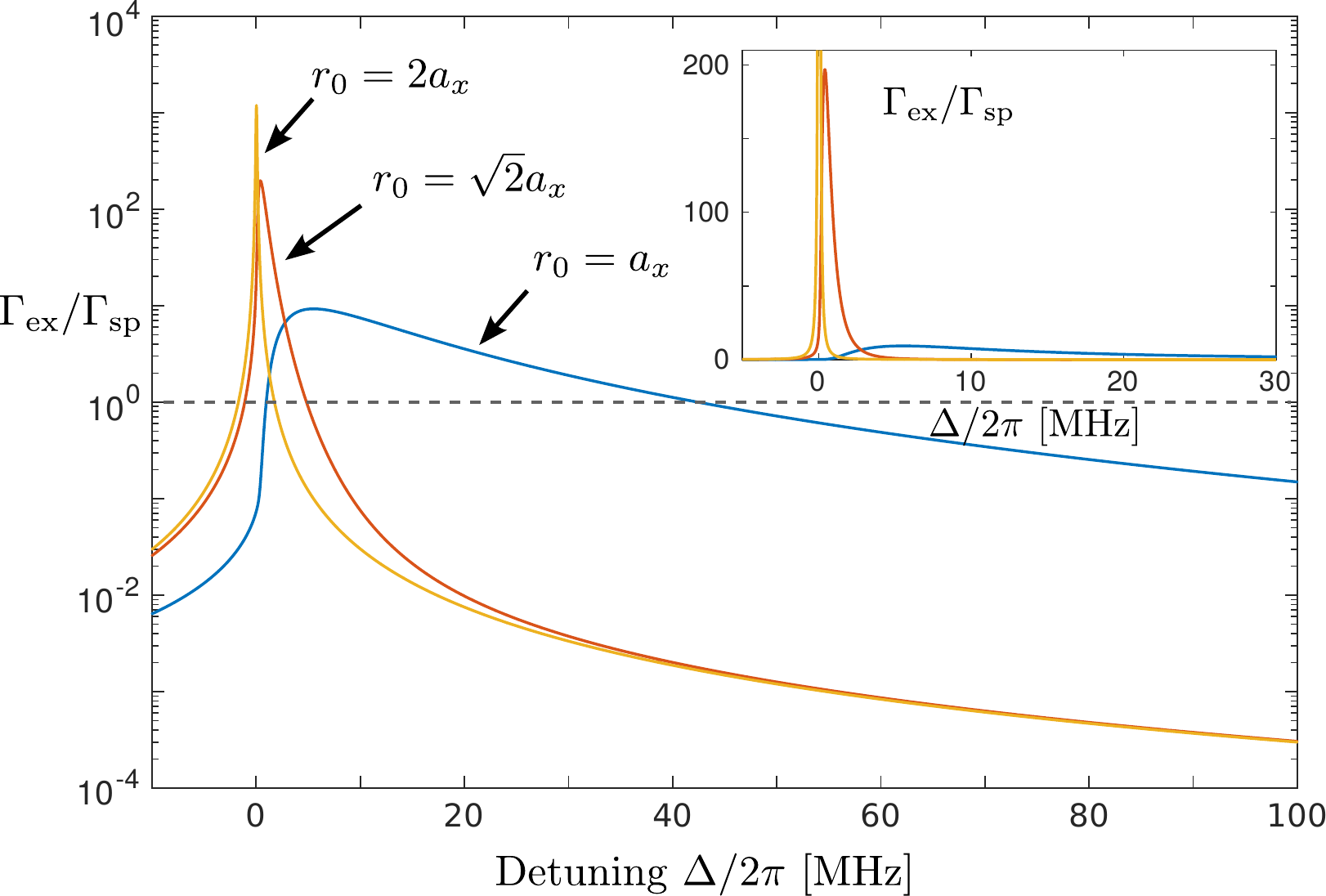, width=.45\textwidth}
\caption{ Excitation rate $\Gamma_\mathrm{ex}/\Gs$ for three different inter particle distances $r_0$ using the parameters from the experiment and $\Omega/2\pi = \SI{500}{\kHz}$. The excitation rate for $r_0 = 2 a_x$ already agrees well with the uncorrelated excitation rate $r_0 \rightarrow \infty$. The inset shows the same rates with linear scale. }
\label{fig:apC-Fig1}
\end{figure}

Here, we discuss an inhomogeneous broadening mechanism resulting from the finite width $\sigma$ of a localized wave packet and the steep slope of an interaction potential $V(r) = C_\alpha/r^\alpha$. The excitation rate $\Gamma_\mathrm{ex}$ for a ground state atom in the presence of an atom already excited to the Rydberg state is strongly suppressed and broadened. As already discussed in \citep{Li2013, Macri2014}, this stems from the coupling to motional states. 

\begin{figure}[h]
\centering
\epsfig{file=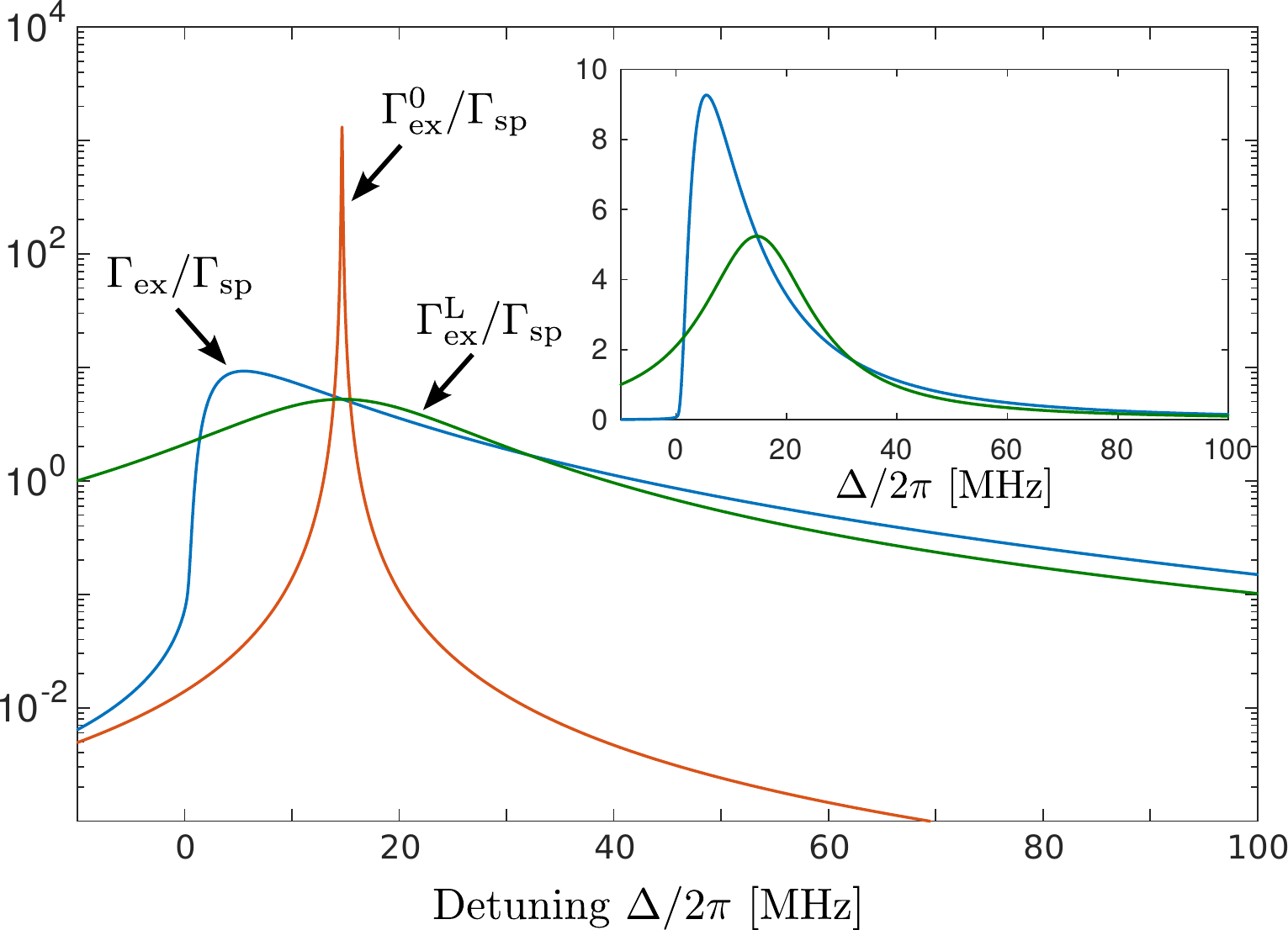, width=.45\textwidth}
\caption{ Comparison of the bare excitation rate $\Gamma_\mathrm{ex}^0/\Gs$ without additional broadening mechanism, the full excitation rate $\Gamma_\mathrm{ex}/\Gs$ eq. \eqref{eq:FullExcitationRate} and the approximate excitation rate $\Gamma_\mathrm{ex}^\mathrm{L}/\Gs$. Here, we use the parameters from the experiment and the inter particle distance is $r_0 = a_x$ and Rabi frequency $\Omega/2\pi = \SI{500}{\kHz}$. The inset shows the same rates with linear scale. }
\label{fig:apC-Fig2}
\end{figure}

Firstly, using the rate equation approximation \cite{Ates2007a}, the bare excitation rate of an atom at distance $r_0$ to an excited Rydberg state is
\begin{equation}
\label{eq:BareExcitationRate}
\Gamma_\mathrm{ex}^0(r_0) = \frac{2\Omega^2\gamma_0}{\gamma_0^2+\left( \Delta - V(r_0) \right)^2}.
\end{equation}
However, this is only true in the limit of vanishing width $\sigma \rightarrow 0$. Now, consider an atom localized in an optical lattice. Using the harmonic oscillator approximation, we can approximate the extent of the localized Wannier wavefunction 
\begin{equation}
\Psi(r) = \left( \frac{1}{\pi \sigma^2} \right)^{1/4} \exp \left(-\frac{r^2}{2\sigma^2} \right)
\end{equation}
with the harmonic oscillator width $\sigma$. In a semi-classical approximation, we interpret $|\Psi(r)|^2$ as the probability distribution of finding an atom at position $r$. Then, the total excitation rate $\Gamma_\mathrm{ex}$ for an atom localized with width $\sigma$ is
\begin{equation}
\label{eq:FullExcitationRate}
\Gamma_\mathrm{ex} = \int_{-\infty}^\infty  \mathrm{d}r \  |\Psi(r)|^2 \ \Gamma_\mathrm{ex}^0(r-r_0),
\end{equation}
where $r_0$ is the average distance to the already excited Rydberg atom. Note, in general we have to evaluate here a 3D integral over the positions of all excited atoms. In this case the effective detuning $\Delta - \sum_j \frac{C_\alpha}{|\vec{r}_j|^\alpha}$ describes the interaction shift landscape generated from all excited Rydberg atoms.

In Fig. \ref{fig:apC-Fig1} we plot the excitation rate $\Gamma_\mathrm{ex}$ for three different experimental relevant distances $r_0$. For two neighboring atoms in a lattice ($r_0 = a_x$), the excitation rate is strongly suppressed and broadened compared to the uncorrelated excitation rate $\Gamma_\mathrm{ex}$ already occurring at $r_0 = 2 a_x$. However, for excitation rates $\Gamma_\mathrm{ex}/\Gs \gg 1$ larger than the lifetime of a single Rydberg excitation, we can still expect cascaded excitations. Here, we identify a large region up to $\Delta/2\pi \simeq \SI{40}{\MHz}$ where an excitation cascade can be possible. 

The exact numerical evaluation of the excitation rate \eqref{eq:FullExcitationRate} in a large scale 3D system using a stochastic many-body rate equation model is numerically challenging. However, to include the strong impact of the inhomogeneous broadening, we use an approximate decoherence rate 
\begin{equation}
\label{eq:ApproximateDecoherenceRate}
\gamma  = \gamma_0 + |\partial_r V(r=r_0)|  \sigma/\sqrt{\pi},
\end{equation}
which can be easily extended to many atoms, see Sec. \ref{sec:Simulation}. The approximate decoherence rate is motivated by the reduced excitation rate $\Gamma_\mathrm{ex} = 2\Omega^2/\gamma$ at $\Delta = V(r_0)$ \cite{Li2013} embedded in a Lorentzian excitation profile $\Gamma_\mathrm{ex}^\mathrm{L} =  2\Omega^2\gamma/(\gamma^2+(\Delta-V(r))^2)$. This decoherence rate includes the energy width of the linearized interaction potential over the width $\sigma$.

In Fig. \ref{fig:apC-Fig2}, we compare the full excitation rate (\ref{eq:FullExcitationRate}) with the excitation rate $\Gamma_\mathrm{ex}^\mathrm{L}$ using the decoherence rate (\ref{eq:ApproximateDecoherenceRate}) for $r_0 =  a_x$. Moreover, we plot the bare excitation rate $\Gamma_\mathrm{ex}^0$ without additional broadening. The approximate rate $\Gamma_\mathrm{ex}^\mathrm{L}$ nicely describes the suppression of excitation and the broadening compared to the bare rate $\Gamma_\mathrm{ex}^0$. In particular, the regime $\Delta/2\pi \gtrsim \SI{18}{\MHz}$ agrees well with the full model \eqref{eq:FullExcitationRate}. However, for smaller detunings $\Delta/2\pi \lesssim \SI{18}{\MHz}$ the simple approximate model is not valid anymore. Therefore, we believe that the approximate decoherence rate is an efficient and good approximation at least for detunings $\Delta/2\pi \gtrsim \SI{18}{\MHz}$.

\section{Hysterese Relaxation}
\label{ap:E}

\begin{figure}[]
\centering
\epsfig{file=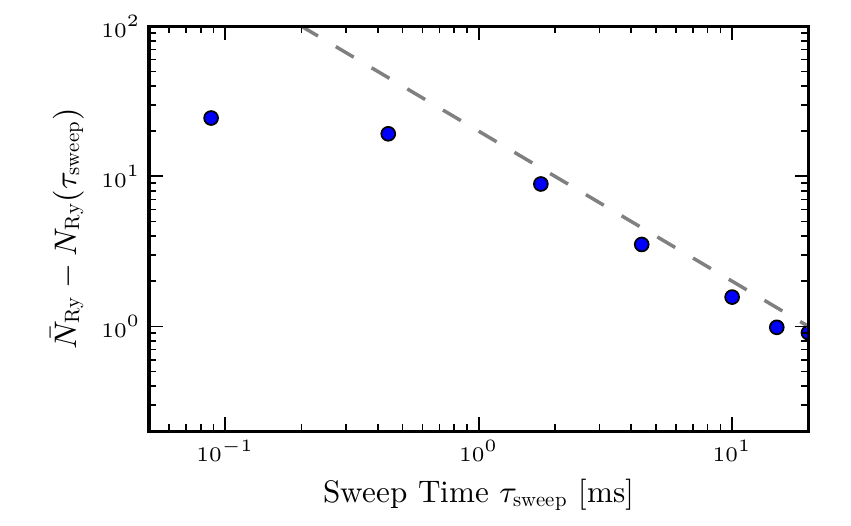, width=.45\textwidth}
\caption{Mismatch between the equilibrium Rydberg excitation number  $\bar{N}_\mathrm{Ry}$ and the Rydberg excitation number $N_\mathrm{Ry}(\tau_\mathrm{sweep})$ after a finite sweep time $\tau_\mathrm{sweep}$. We use the same parameters as in Fig. \ref{fig:SizeScaling-distribution}b. The hysteresis scan starts at $\Delta/2\pi = \SI{40}{\MHz}$ and stops at $\Delta/2\pi = \SI{10}{\MHz}$. The dashed line indicates a scaling $\propto \tau_\mathrm{sweep}^{-1}$. }
\label{fig:apE-HystereseRelaxation}
\end{figure}

Here, we discuss a simple model for the relaxation of the hysteresis area. In the case of a single quench from a parameter $\Delta_0$ to $\Delta_1$, we typically have an exponential relaxation of some observable $n$ with relaxation timescale $T_1$:
\begin{equation}
\label{eq:SingleQuench}
n(t) = \bar{n}_1 + (n_0-\bar{n}_1) e^{-t/T_1}.
\end{equation}
Here, $\bar{n}_1$ is the equilibrium value of the observed parameter and $n_0$ its initial value, which is not necessarily at equilibrium. 

In the case of a hysteresis, we perform $N$ small quenches $\Delta_j$ to $\Delta_{j+1}$ on a timescale $\tau$. The total sweep time is $\tau_\mathrm{sweep} = N \tau$. We denote $\bar{n}_j$ the equilibrium value at $\Delta_j$ and $n_j = n(j\tau)$ the non equilibrium value after time $j\tau$. The relaxation time $T_j = T(\Delta_j)$ might depend on the tuning parameter $\Delta_j$. Then, after $N$ consecutive quenches of the form eq. \eqref{eq:SingleQuench}, the non equilibrium value of the observable $n$ is given by 
\begin{align}
\label{eq:DiscreteQuenches}
n_N =& n_0 \exp \left( -\sum_{j=1}^N \frac{\tau}{T_j} \right) \nonumber \\
 &+ \sum_{j=1}^N \bar{n}_j \left( 1-e^{-\tau/T_j} \right) \exp \left( - \sum_{k=j}^{N-1} \frac{\tau}{T_k} \right).
\end{align}

Now, let us perform a continuum limit, where we keep the sweep rate $\dot{\Delta} = \frac{\Delta_N-\Delta_0}{\tau_\mathrm{sweep}}$ constant and increase the number of steps $N \rightarrow \infty$. We assume equidistant steps in the quench parameter $\delta = \Delta_{j+1}-\Delta_j$. Since the sweep time $\tau_\mathrm{sweep}$ and the parameter regime $\Delta_N-\Delta_0$ are constant, we have $\delta,\tau \rightarrow 0$. This allows us to approximate the exponential
\begin{equation}
1-e^{-\tau/T_j} \simeq \tau/T_j
\end{equation}
and replace the sum with an integral. We obtain the continuum counterpart of eq. \eqref{eq:DiscreteQuenches}:
\begin{align}
n(\tau_\mathrm{sweep}) = n_0 \exp \left( -\frac{1}{\dot{\Delta}} \int_{\Delta_0}^{\Delta_N} \mathrm{d}\Delta \ \frac{1}{T(\Delta)} \right)
\nonumber \\
+\frac{1}{\dot{\Delta}} \int_{\Delta_0}^{\Delta_N} \mathrm{d}\Delta \left[ \frac{\bar{n}(\Delta)}{T(\Delta)} \exp \left( -\frac{1}{\dot{\Delta}} \int_{\Delta}^{\Delta_N} \mathrm{d}\tilde \Delta \ \frac{1}{T(\tilde \Delta)} \right) \right].
\end{align}
For a constant relaxation time $T = T(\Delta)$, we obtain eq. \eqref{eq:HystereseRelaxation}. 

In Fig. \ref{fig:apE-HystereseRelaxation} we show the relaxation of the Rydberg excitation number with increasing sweep time $\tau_\mathrm{sweep}$. The result shows the expected power law relaxation $\tau_\mathrm{sweep}^{-1}$.

\bibliography{ClusterDynamics}

\begin{thebibliography}{52}
\bibitem [{foo()}]{footnote_decay}%
  \BibitemOpen
  \href@noop {} {\bibinfo  {journal} {The assumption of a steady state breaks
  down, whenever the timescale for the loss of atoms becomes of similar size as
  the correlation time $\tau^{(2)}$. In our analysis, we discard such data
  points.}\ }\BibitemShut {NoStop}%
\end{thebibliography}
\end{document}